\documentclass[journal, onecolumn,12pt, draftclsnofoot]{IEEEtran}
\usepackage{citesort}
\usepackage{amsmath,amsthm}
\usepackage{algorithmic,algorithm}
\usepackage{array}
\usepackage{amsfonts}
\usepackage{graphicx,color,overpic,psfrag,epsfig}
\usepackage{bm}
\usepackage{footmisc}
\usepackage{xcolor}
\usepackage{color}
\usepackage{subfigure}
\usepackage{epstopdf}
\usepackage{amssymb}
\newcommand{\argmin}{\operatornamewithlimits{arg\, min}}
\newcommand{\argmax}{\operatornamewithlimits{arg\, max}}

\newcommand{\qD}{{\bf D}}

\newcommand{\qc}{{\bf c}}
\newcommand{\qw}{{\bf w}}

\newcommand{\qq}{{\bf q}}

\newcommand{\qn}{{\bf n}}
\newcommand{\qx}{{\bf x}}
\newcommand{\qv}{{\bf v}}

\newcommand{\calA}{{\mathcal{A}}}
\newcommand{\calC}{{\mathcal{C}}}

\newcommand{\calP}{{\mathcal{P}}}
\newcommand{\calK}{{\mathcal{K}}}
\newcommand{\calS}{{\mathcal{S}}}
\newcommand{\calQ}{{\mathcal{Q}}}
\newcommand{\calW}{{\mathcal{W}}}

\newtheorem{theorem}{Theorem}

\newtheorem{lemma}{Lemma}
\newtheorem{assumption}{Assumption}
\newcommand{\Ex}{{\mathbb{E} }}
\begin{document}

\title{Multi-Armed Bandit Based Client Scheduling for Federated Learning} 


\author{Wenchao Xia, Tony Q. S. Quek,~\IEEEmembership{Fellow,~IEEE},  Kun Guo, Wanli Wen, \\ Howard H. Yang,~\IEEEmembership{Member,~IEEE}, and Hongbo Zhu
\thanks{W. Xia, T. Q. S. Quek,  K. Guo, W. Wen, and H. H. Yang  are with the  Information  Systems  Technology  and  Design Pillar, Singapore University of Technology and Design, Singapore 487372, (e-mails: wenchao\_xia@sutd.edu.sg, tonyquek@sutd.edu.sg,  kun\_guo@sutd.edu.sg, wanli\_wen@sutd.edu.sg, howard\_yang@sutd.edu.sg).}
\thanks{H. Zhu is with the Jiangsu Key Laboratory of Wireless Communications, and also with the Engineering Research Center of Health  Service System Based on Ubiquitous  Wireless  Networks, Ministry of Education,  Nanjing University of Posts and Telecommunications, Nanjing 210003, China, (e-mail: zhuhb@njupt.edu.cn).}
\vspace{-1.6cm}
}
\maketitle
\begin{abstract}\footnotesize
By exploiting the computing power and local data of  distributed clients, federated learning (FL) features ubiquitous properties such as reduction of communication overhead and preserving data privacy. In each communication round of FL, the clients update local models based on their own data and  upload their local updates via wireless channels.
However, latency  caused by hundreds to thousands of communication rounds remains a bottleneck in FL.  To minimize the training latency, this work provides a multi-armed bandit-based framework for online client scheduling (CS) in FL  without knowing wireless channel state information and  statistical characteristics of clients.
Firstly, we  propose a CS algorithm based on the upper  confidence bound  policy (CS-UCB) for ideal scenarios where local datasets of clients are independent and identically distributed (i.i.d.) and balanced. An upper bound of the expected performance regret of the proposed CS-UCB algorithm is provided, which indicates that the regret grows  logarithmically over communication rounds. Then, to address  non-ideal scenarios with non-i.i.d. and unbalanced properties of local datasets and  varying availability of clients, we further propose a CS algorithm  based on the UCB policy  and virtual queue technique (CS-UCB-Q). An upper bound is also derived, which shows that the expected performance regret of the proposed CS-UCB-Q algorithm can have a sub-linear growth over communication rounds under certain conditions. Besides, the convergence performance of FL training is also analyzed. Finally, simulation results validate the efficiency of the proposed algorithms.
\end{abstract}
\begin{IEEEkeywords}
Federated learning, client scheduling, client selection, multi-armed bandit (MAB)
\end{IEEEkeywords}
\section{Introduction}
The unprecedented amount of data and the increasing number of mobile devices, e.g., smart phones, tablets, and Internet of Things (IoT) devices, in wireless communication networks, together with the recent breakthroughs in machine learning (ML),  are revolutionizing our way of life. While traditional ML frameworks rely on the availability of a large amount of data  in a centralized  computing entity, this is not always feasible in IoT applications due to: (a) the reluctance of sharing private data from the end users, (b) the growing security concerns, and (c) the large communication overhead required to transmit raw data to centralized ML processors. In response, federated learning (FL) \cite{mcmahan2016communication} was recently proposed to leverage massively distributed computing power, e.g., on IoT devices or smart phones, to collaboratively train a shared ML model without direct access to  raw data. These devices, which are referred to as clients,  train an ML model on their local datasets, respectively, and  then upload their local model parameters (e.g., model weights or gradients) to a remote server for model aggregation. Thus, the massive amount of distributed and privacy-sensitive data on the clients can be well exploited without leak of privacy.

Unfortunately, latency incurred by wireless transmission and local computation on clients  remains a bottleneck in FL, since hundreds to thousands of communication rounds are required to reach a desired model accuracy, especially when the number of participating clients in the  training process of FL is large \cite{bonawitz2019towards}. Specifically,  the large propagation latency caused by dynamic wireless environment and the large computation delay  due to limited computing power of clients will degrade model performance, if with a limited training time budget. The approaches to reducing the training latency in FL proposed in existing works can be roughly divided into three main categories: update compression, over-the-air computation, and communication reduction.
Belonging to the first category,  gradient quantization \cite{du2019high} and gradient sparsification  \cite{aji2017sparse,abad2019hierarchical} can be used to improve the communication efficiency. Besides, two ways to reduce the uplink communication costs: structured updates and sketched updates of model parameters were proposed in \cite{lin2017deep}. Recognizing the superposition property of wireless multiple-access channels, computation-over-the-air approaches which harness interference instead of suppressing interference are shown to be able to accelerate global model aggregation,  such as broadband analog aggregation adopted in \cite{yang2018federated,zhubroad2019band,amiri2019machine}. The last category has also been investigated in some works. For example, the adaptive communication strategies with dynamic local updates and with dynamic model aggregation were proposed by \cite{wang2019adaptive} and \cite{kamp2018efficient}, respectively.

In addition to the aforementioned methods, client scheduling (CS) is also  an important direction to reduce training latency in FL, especially in the scenarios with massive number of clients but limited wireless channels. Different scheduling criterions lead to different performance. For example, in \cite{chen2016revisiting}  only a portion of clients with fast response were chosen for aggregation and the stragglers with slow response were temporarily dropped in a certain communication round, in order to avoid a long wait for the stragglers. \cite{nishio2019client,yang2018federated} maximized the number of selected clients in each communication round.  \cite{yang2019age} proposed a scheduling policy by jointly accounting for the staleness of the received parameters and the instantaneous channel qualities. Another scheduling criterion is the update significance such as  model variance \cite{kamp2018efficient} and gradient variance \cite{chen2018lag}. Besides, the work in \cite{chen2019joint,zeng2019energy} considered  a joint problem of wireless resource allocation and CS. Note that these works performed CS based on  the assumption that some prior information is available, such as wireless channel state information (CSI) and  computing resource usage of clients. However, it is not easy to have access to such information  when the number of clients is greatly large. Secondly, most of these works, except for \cite{ren2019accelerating,shi2019device}, only investigated the transmission time consumption of uploading local updates and distributing global model parameters, but did not take into account the computation time consumption of local training on clients. In addition, due to the  potential ``deep fade" of wireless channels or the exhausted computing power, some clients may lose their connectivity and be unavailable temporarily.  Thus, the availability of clients during FL training is also a practical challenge, but was not considered in these existing works.  Motivated by these facts,  a new CS policy in FL training is proposed in this paper to deal with these practical challenges.

In this work, given the number of communication rounds required to achieve a certain level of test accuracy \cite{ma2017distributed,dinh2019federated}, we aim to minimize the total wall-clock time consumption including transmission time and local computation time. We consider a system where  an access point (AP) equipped with a central processor serves a large amount of clients.  However, only a limited number of wireless channels are available, i.e., the number of wireless channels is much smaller than that of clients. Thus, scheduling the appropriate clients  to access to the wireless channels for parameter updating during different communication rounds is of necessity. We study the CS problem in FL in two scenarios: ideal and non-ideal scenarios. In the ideal scenario, the clients are always available and the dataset of each client is balanced and independent and identically distributed (i.i.d.). In the non-ideal scenario, the clients are not always available during the FL training process and the datasets of different clients are unbalanced and non-i.i.d.  Due to the lack of prior knowledge of training time of the clients, we need to learn the related statistical information while performing CS, so as to help make better decisions in the future. Thanks to the development of reinforcement learning technique, we reformulate the CS problem as a multi-armed bandit  (MAB) program, which is a powerful tool for scheduling and resource allocation problems \cite{liu2010distributed,Ali2020sleeping,bonnefoi2017multi}. To our best knowledge, this work is the first attempt to apply the MAB tool to the CS problem in FL.


The contributions of this work are summarized as follows:

\begin{itemize}
  \item We formulate a CS problem which aims to minimize the  time consumption of the whole FL training including transmission time and local computation time in both ideal and non-ideal scenarios. Then  we provide an MAB-based framework to learn to schedule clients online in FL training without knowing wireless CSI and dynamics of computing resource usage of clients.
  \item For the ideal scenario, we propose a CS algorithm based on  upper bound confidence (UCB) policy  \cite{auer2002finite}, namely CS-UCB algorithm, which strikes a balance between the exploitation of actions that performed well in the past and the exploration of actions that might return higher rewards in the future. An upper bound on regret is provided and shows the expected performance regret grows in the logarithmic way over communication rounds.
  \item For the non-ideal scenario with non-i.i.d. and unbalanced properties of local datasets and  dynamic availability of clients, we introduce the fairness constraint to ensure each client can participate in a certain proportion of the communication rounds during the training process. We further propose a CS algorithm  based on UCB policy and virtual queue technique, namely CS-UCB-Q algorithm. An upper bound on regret is also derived and shows that the expected performance regret of the proposed CS-UCB-Q algorithm can have a sub-linear growth over communication rounds under certain conditions.
  \item The convergence performance in both the ideal and non-ideal scenarios is analyzed. It is found that increasing the number of the participating clients in each round can improve the convergence rate in the ideal scenario, but  the convergence performance has a weak dependence on the number of the participating clients in each round in the non-ideal scenario.
  \item Simulation results validate the efficiency of the proposed CS-UCB and CS-UCB-Q algorithms. Besides, A tradeoff between the performance regret and the speed of convergence to a point where the fairness constraint is  satisfied is revealed.
\end{itemize}

The remainder of this paper is organized as follows. Section \ref{section system model} introduces the system model and formulates the CS problems in both the ideal and non-ideal scenarios.  Sections \ref{section scheduling in IS} and \ref{section scheduling in NS}  propose the CS-UCB and  CS-UCB-Q algorithms for the ideal and non-ideal scenarios, respectively, and an upper bound on regret is derived for each of the proposed algorithms.  In addition, the convergence performance is also analyzed. Numerical results are presented in Section \ref{section numerical results}.  Finally, conclusion is drawn in Section \ref{section conclusions}.

\vspace{-0.5cm}
\section{System Model and Problem Formulation}\label{section system model}
We consider a wireless network which is composed of a single-antenna AP and massive single-antenna clients (e.g., smart phones, sensors, and cameras). The set of clients is denoted as $\calK$ with the cardinality $|\calK|=K$.
Each client $k\in\calK$ has a local dataset $\calW_k$ consisting of $s_k=|\calW_k|$ sample points, based on which local training is performed.  The FL training aims to minimize the weighted global loss function \cite{haddadpour2019convergence,li2019convergence}, which is given as
\begin{equation}
   \min_{\qx} G(\qx)=\sum_{k\in\calK}\varpi_k G_{k}(\qx),
\end{equation}
where $\varpi_k>0$ is the weight of client $k$ with $\sum_{k\in\calK}\varpi_k=1$ and   $G_{k}(\qx)$ is the local empirical loss of client $k$, which is defined as
\begin{equation}
  G_{k}(\qx)=\frac{1}{s_k}\sum^{s_k}_{s^{\prime}=1}g(\qx;\qw_{ks^{\prime}}),
\end{equation}
where $\qw_{ks^{\prime}}$ is the $s^{\prime}$-th sample point in $\calW_k$ and $g(\qx;\qw)$ is a loss function of a parameter vector $\qx$ associated with the sample point $\qw$.

We assume a fixed amount of spectrum is available and  equally divided into $N$ orthogonal  radio access channels. A wireless channel is allocated to at most one client in each round such that there is no inter-client interference.   Due to the scarce of the  spectrum  resource, the number of available channels is much smaller than that of the candidate clients, i.e., $N<K$.
The training procedure of FL is an iterative process consisting of a number of communication rounds.  In each round $t$, the AP chooses a subset $\calS(t)\subseteq\calK$ of the clients  and then \textbf{distributes} the weight vector $\qx(t)$ of the global model to the selected clients. After receiving the global model weights,  each of the selected clients individually \textbf{updates} the global model by computing the gradients of their local loss functions  based on their own private data and then \textbf{uploads} the updated gradients  to the AP for model \textbf{aggregation}, i.e.,
\begin{equation}\label{aggregation}
  \qx(t+1)=\qx(t)-\gamma\qv(t),
\end{equation}
where  $\qv(t)=\sum_{k\in\calS(t)}\frac{\varpi_k}{\sum_{k^{\prime}\in\calS(t)}\varpi_{k^{\prime}}} \nabla G_{k}(\qx(t))$.
\vspace{-0.5cm}
\subsection{Training Latency}
Since  the AP has abundant  computational resource  compared to the clients, the latency incurred by global model aggregation is negligible. According to the above analysis,  the time cost of each round $t$ of client $k$ depends on three main components: distribution time, local update time, and upload time, denoted by $\tau^{\textrm{D}}_k(t)$, $\tau^{\textrm{LU}}_k(t)$,  and  $\tau^{\textrm{U}}_k(t)$, respectively. Then, the total time consumed by client $k$ in round $t$ is given  as
\begin{equation}\label{round_time_client}
  \tau_k(t)= \min\{\tau^{\textrm{D}}_k(t)+\tau^{\textrm{U}}_k(t)+\tau^{\textrm{LU}}_k(t),\tau_{\max}\},
\end{equation}
where $\tau_{\max}$ represents the maximal interval  of each communication round, which is used to avoid an endless wait caused by possible stragglers. The distribution time $\tau^{\textrm{D}}_k(t)$ and upload time $\tau^{\textrm{U}}_k(t)$  of client $k$ in  round $t$ relies on the size  of the model parameters and the wireless channel between the AP and  client $k$.
It is worth  noting that the available computing resource   on each client varies over time, because a client can execute multiple processing tasks at the same time. For example, a smart phone is used to play games, while performing the local training. To capture the dynamics of local computing power of the clients in FL training and for simplicity, we assume the amount  of  available computing power on each client $k$ in round $t$ for the local training, denoted by $\phi_k(t)$ in sample points per second, is  i.i.d. over time according to a certain distribution and its expectation is also unknown a priori. Because of the heterogeneity of available computing power, the resulting local update time $\tau^{\textrm{LU}}_k(t)$ is also different among the clients.

Fortunately, instead of obtaining the values of the three main components separately, the AP in our work observes $\tau_k(t)$ directly, which significantly reduces difficulties and overhead.
\vspace{-0.5cm}
\subsection{Availability Constraint}
During the FL process, some clients can be unavailable temporarily due to, for example, the  poor channel conditions or the exhausted computing power. We define a binary variable $a_k(t)$ and  if  client  $k$ is available in round $t$, then $a_k(t)=1$; otherwise $a_k(t)=0$.
We further define  $\calA(t)\triangleq\{k\in\mathcal{K}|a_k(t)=1\}\in \calP(\calK) $ as the set of the available clients  in round $t$  where $\calP(\calK)$ is the power set of $\calK$. For simplicity, it is assumed that the set of the available clients is i.i.d.\footnote{For the cases where the set of the available clients is non-i.i.d. over time, we can introduce a Markov process to analyze the upper bound on performance regret \cite{ahmad2009multi,whittle1988restless}.} over time and the corresponding distribution of the available clients, $\hat{P}_{\calA}(e)=\hat{P}(\mathcal{A}(t)=e), e\in\calP(\calK)$, is unknown in advance.  However,  $\mathcal{A}(t)$  is revealed to the AP at the beginning of each round $t$.
Note that in the considered network the number of the clients is much larger than that of the available channels, thus the AP has to choose a subset $\calS(t)$ from the  available clients, i.e.,
\begin{equation}\label{availability constraint}
  \calS(t)\triangleq \{\calS(t) \subseteq \calA(t):|\calS(t)|\leq\min\{N, |\calA(t)| \}\}\in \calQ(\calA(t)),
\end{equation}
where $|\mathcal{S}(t)|$ represents the cardinality of  $\mathcal{S}(t)$ and $\calQ(\calA(t))$ is the power set of $\calA(t)$.
\vspace{-0.5cm}
\subsection{Fairness Constraint}
Besides the availability constraint, the non-i.i.d. and unbalanced properties of local datasets of the clients  should be addressed.  More specifically, the local dataset of any particular client cannot represent the distribution of the whole, and the sizes of the local datasets of different clients are different, because different clients have various activity characteristics. An effective solution to the considered issue is Federated Averaging approach \cite{mcmahan2016communication} in which  a subset of the clients are enrolled in the FL process in each round to cooperatively train a shared model.
However, the importance of the local datasets of different clients is different. Generally speaking, the local dataset which has a larger size and  whose distribution is  more similar to the global distribution plays a more important role, and the corresponding clients  should participate in more communication rounds.
Thus, fairness constraint is introduced to ``tell" each client that how many communication rounds they should participate in.  The fairness constraint not only can  avoid the occurrence of abandoning the slow but important clients due to the pursuit of low delay, but also can make the important clients be involved in more rounds by setting a large fairness factor. A binary variable $b_k(t)$  is defined as an indicator with $b_k(t)=1$ indicating that client $k$ is selected in round $t$, and $b_k(t)=0$ otherwise. Then, the fairness constraint is formulated as \cite{li2019combinatorial}
\begin{equation}\label{fairness constraint}
\setlength{\abovedisplayskip}{3pt}
\setlength{\belowdisplayskip}{3pt}
  \lim\limits_{T\rightarrow\infty}\inf\frac{1}{T}\sum\nolimits_{t=1}^{T}\Ex[b_k(t)]\geq c_k, \forall k\in\mathcal{K},
\end{equation}
where $\Ex[\cdot]$ is the expectation operator and  $c_k\in[0,1)$ is the minimum fraction of communication rounds required to choose client $k$. Since the  distribution of the local dataset is usually unknown, we use the size of the local dataset as a metric to roughly describe the importance and set the fairness constraint. Here, we provide a heuristic example of the setting of the fairness constraint. We first find client $\hat{k}=\argmin_{k\in\calK}{s_k}$ and set $c_{\hat{k}}=c_{\min}$, where $\varphi\frac{Ns_{\hat{k}}}{\sum_{k\in\calK}s_k}\leq c_{\min}\leq\frac{Ns_{\hat{k}}}{\sum_{k\in\calK}s_k}$ and $0\leq\varphi<1$. The fairness constraint values of the other clients are set as  $c_k=\frac{s_k}{s_{\hat{k}}}c_{\min},  k\neq \hat{k}$.

These constraint values, $\{c_k\}^K_{k=1}$,  are incorporated into a vector $\mathbf{c}=[c_1,c_2,\ldots,c_K]^\dagger$, where $[\cdot]^\dagger$ denotes the transpose operator.  If there exists a policy that can find a time sequence $\{\calS(t), t\geq 1\}$  to satisfy the  constraint in \eqref{fairness constraint}, then the constraint vector $\mathbf{c}$ is said to be feasible, otherwise infeasible. For example, any $\mathbf{c}$ satisfying $\sum_{k\in\calK}c_k>N$ is infeasible. Furthermore, we denote by  $\calC$  the maximal feasibility region which includes all such  feasible $\mathbf{c}$'s.

\vspace{-0.5cm}
\subsection{Problem Formulation}
During each communication round, the AP cannot perform model aggregation until all the selected clients finish data uploading, thus the time consumption of each  round $t$ is determined by the slowest client, i.e.,
\begin{equation}
\setlength{\abovedisplayskip}{3pt}
\setlength{\belowdisplayskip}{3pt}
 \hat{\tau}(t,\calS(t))=\max_{k\in \calS(t)}\tau_k(t).
\end{equation}
In the following, for simplicity, $\hat{\tau}(t,\calS(t))$ is written as  $\hat{\tau}(\calS(t))$  without ambiguity. This work considers both ideal and non-ideal scenarios. \textbf{In the ideal scenario}, all the clients are available anytime and  the local datasets of the clients  are i.i.d. and balanced,  thus it is unnecessary to take the availability constraint and the fairness constraint into account. In addition, the convergence rate improves substantially as the number of the participating clients in each round increases \cite{stich2018local}.  Thus, constraint \eqref{availability constraint} can be simplified as
 \begin{equation}\label{availability constraint2}
  \calS(t)\triangleq \{\calS(t): |\calS(t)|=N\}\in \calP(\calK).
\end{equation}
We aim to minimize the time cost of the FL training process, by finding  a decision sequence $\{\calS(t), t\geq 1\}$, i.e.,
\begin{equation}\label{original pro}
\setlength{\abovedisplayskip}{3pt}
\setlength{\belowdisplayskip}{3pt}
\begin{split}
  \mathop{\min}_{\{\calS(t),t\geq 1\}} \ \sum_{t=1}^{T} \bar{\tau}(\calS(t))\eta(t), \ \ \text{s.t.} \  \eta(t)\in\{0,1\} \ \ \text{and} \ \ \eqref{availability constraint2},
\end{split}
\end{equation}
where $T$ is a large constant,  $\bar{\tau}(\calS(t))=\frac{\hat{\tau}(\calS(t))}{\tau_{\max}}\in(0,1]$ is the normalized delay, and $\eta(t)=1$ is an indicator, which equals 1 if the FL process  does not converge and 0 otherwise \cite{chen2020convergence}.

However, \textbf{in the non-ideal scenario}, the clients can be unavailable during the learning process and the local datasets of different clients are non-i.i.d. and unbalanced, thus the fairness constraint and the availability constraint should be addressed. Note that under the non-i.i.d. scenario, the convergence rate has a weak dependence on the number $|\calS(t)|$ of the participating clients in each round \cite{li2019convergence}.  As mentioned in Section II,  the number of the clients is much larger than that of the available channels in the considered network. Thus, for simplicity, it is reasonable to  make full use of available channels and select as many clients as possible in each round, i.e.,
\begin{equation}\label{availability constraint3}
\setlength{\abovedisplayskip}{3pt}
\setlength{\belowdisplayskip}{3pt}
  \calS(t)\triangleq \{\calS(t) \subseteq \calA(t):|\calS(t)|=\min\{N, |\calA(t)| \}\}\in \calQ(\calA(t)).
\end{equation}
Then, the underlaying problem, with the availability and fairness constraints,  is formulated as
\begin{equation}\label{original problem}
\setlength{\abovedisplayskip}{3pt}
\setlength{\belowdisplayskip}{3pt}
\begin{split}
  \mathop{\min}_{\{\calS(t),t\geq 1\}} \ \sum_{t=1}^{T} \bar{\tau}(\calS(t))\eta(t),  \ \ \text{s.t.} \ \eta(t)\in\{0,1\}, \eqref{fairness constraint}, \ \text{and} \ \eqref{availability constraint3}.
\end{split}
\end{equation}
Solving the scheduling problems \eqref{original pro} and \eqref{original problem} is not easy because of the lack of prior information. We do not know  either the dynamics of computing power of the clients or wireless CSI between the AP and clients, so that we cannot make the optimal decisions. In order to address this issue, we must learn some important information, i.e., the statistical information about time consumption of each client, from the feedback from the previous decisions. Besides, the fairness constraint and the availability constraint  increase the difficulty in solving problem~\eqref{original problem}. In the following, we will first handle problem~\eqref{original pro}, and then we take a further step to deal with  problem~\eqref{original problem}.
\vspace{-0.5cm}
\section{Scheduling in Ideal Scenario}\label{section scheduling in IS}
In this section, we first give a short introduction to MAB problems, then reformulate problem \eqref{original pro} into an MAB problem and propose an algorithm to find the decision sequence $\{\calS(t)\}^T_{t=1}$ in the ideal scenario. Finally, we  provide some theoretical analysis on the performance of the proposed algorithm.
\vspace{-0.5cm}
\subsection{MAB Problems}
An MAB problem is one of sequential decision problems, where a player should make a decision, in each time slot, about which arms of the bandits are pulled \cite{lai1985asymptotically}. We refer to arm pulls as an action and  a reward (payoff) is observed when an action is taken. The aim of an MAB problem is to make sequential decisions to maximize the total reward obtained in a sequential of actions. Due to the limited information available, the player has to strike a balance between the exploitation of actions that performed well in the past and the exploration of actions that might return higher rewards in the future \cite{bubeck2012regret}.
\vspace{-0.5cm}
\subsection{Problem Reformulation}
The scheduling problem \eqref{original pro} can be modelled as an MAB problem, where the AP  and the clients can be regarded as the player and the arms, respectively. The chosen subset of  the arms $\calS(t)$ is  referred to as a super arm and its corresponding normalized reward is $r(\calS(t))=-\bar{\tau}(\calS(t))+1\in[0,1)$.  Accordingly, the objective of problem  \eqref{original pro} can be interpreted as finding a time sequence of actions, according to a particular policy, to maximize the cumulative reward, i.e.,
\begin{equation}\label{regret pro}
\max_{\{\calS(t), t\geq 1\}} \sum_{t=1}^{T} r(\calS(t))\eta(t).
\end{equation}

For the scheduling problem \eqref{regret pro},  we evaluate the policies of action selection with respect to \emph{\textbf{regret}} \cite{lai1985asymptotically}, which is defined as the difference between the expected reward of the optimal actions  and that obtained by the given policy. Note that minimizing the regret is equivalent to maximizing the total reward. We denote by $\mu_k=\Ex\left[r_k(t)\right]$ the expectation of the normalized reward $r_k(t)=-\frac{\tau_k(t)}{\tau_{max}}+1$ of arm $k$, and the optimal super arm is
\begin{equation}
  \calS^{\ast}=\argmax_{\calS\subseteq\calK,|\calS|=N}\{\min_{k\in \calS}\mu_k\}.
\end{equation}
Assume that $\eta(T)=1$ and $\eta(T+1)=0$, meaning that the FL training converges at the end of round $T$.  Then, the cumulative regret of a given policy $\pi_1$ is
\begin{equation}\label{regret 1}
  \Sigma^{\pi_1}=T\mu(\calS^{\ast})-\Ex\left[\sum_{t=1}^{T} r\Big(\calS(t)\Big)\right],
\end{equation}
where $\mu(\calS)=\min_{k\in\calS}\mu_k$ is the  expectation of the reward of the super arm $\calS$. Then, problem \eqref{original pro} can be reformulated as
\begin{equation}\label{reward mini1}
\begin{split}
 \mathop{\min}_{\{\calS(t),t\geq 1\}} \ \Sigma^{\pi_1}, \ \ \text{s.t.} \   \eqref{availability constraint2}.
 \end{split}
\end{equation}
\vspace{-0.5cm}
\subsection{Proposed Algorithm}
\setlength{\intextsep}{0pt}
\setlength{\textfloatsep}{0pt}
\begin{algorithm}[htb]\small
\caption{Proposed CS-UCB algorithm for problem  \eqref{reward mini1}.}
\label{alg1}
\begin{algorithmic}[1]
\STATE \textbf{Initialization:} Set $\calK^{\prime}=\calK$ and  $\calK^{\prime\prime}=\emptyset$.
\FOR{$t=1,2,\ldots,\kappa$}
\IF{$|\calK^{\prime}|< N$}
\STATE Randomly choose a set $\calS$ of $\kappa N-K$ arms from $\calK^{\prime\prime}$ and update $\calK^{\prime}=\calK^{\prime}\cup\calS$.
\ENDIF
\STATE  Randomly choose a super arm $\calS(t)$ with $|\calS(t)|=N$ from $\calK^{\prime}$ and update $\calK^{\prime}=\calK^{\prime} \setminus\calS(t)$ and $\calK^{\prime\prime}=\calK^{\prime\prime} \cup\calS(t)$.
\STATE Update $\mathbf{y}(t)$ and $\mathbf{z}(t)$ according to \eqref{update y} and \eqref{update z}, respectively.
\ENDFOR
\STATE \textbf{Main loop:}
\WHILE{$T-t\geq0$}
\STATE $t=t+1$.
\STATE Choose a super arm $\calS(t)$ according to \eqref{arm selection policy}.
\STATE Update $\mathbf{y}(t)$ and $\mathbf{z}(t)$ according to \eqref{update y} and \eqref{update z}, respectively.
\ENDWHILE
\end{algorithmic}
\end{algorithm}

As shown in \textbf{Algorithm \ref{alg1}}, the proposed algorithm for problem \eqref{reward mini1}  is based on  UCB policy \cite{auer2002finite,gai2012combinatorial}, namely CS-UCB algorithm. The key idea behind this algorithm is that we observe the feedbacks for each arm, rather than for each super arm (or each action) as a whole. The same arm can be observed in different actions and we  collect and exploit information about the reward of one certain arm from the operations of different actions, in order to make better decisions in the future.

We define two vectors $\mathbf{y}(t)\in\mathbb{R}^{K \times1}$ and $\mathbf{z}(t)\in\mathbb{R}^{K\times1}$. $y_k(t)$ is the $k$-th element of  $\mathbf{y}(t)$ and represents the sample mean of the observed reward values  of arm $k$ up to the current round $t$, which is updated by the following rule:
\begin{equation}\label{update y}
  y_k(t)=
  \begin{cases}
  \frac{y_k(t-1)z_k(t-1)+r_k(t-1)}{z_k(t-1)+1}, \ & \text{if} \ k\in\calS(t),\\
  y_k(t-1), \ & \text{else},
  \end{cases}
\end{equation}
where $z_k(t)$ is the $k$-th element of $\mathbf{z}(t)$ and denotes the number of times that arm $k$ is played by the end of round $t$, which is updated as follows:
 \begin{equation}\label{update z}
  z_k(t)=
  \begin{cases}
  z_k(t-1)+1, \ & \text{if} \ k\in\calS(t),\\
  z_k(t-1), \ & \text{else}.
  \end{cases}
\end{equation}
Note that both  $\mathbf{z}$ and $\mathbf{y}$ are initialized to $\bm{0}$ when $t=0$, i.e.,  $\mathbf{z}(0)=\mathbf{y}(0)=0$.

In \textbf{Algorithm \ref{alg1}}, we use $\kappa=\lceil\frac{K}{N}\rceil$ rounds for initialization to make sure each arm is played at least one time.  After initialization, \textbf{Algorithm \ref{alg1}} enters the main loop and selects a super arm  in each round $t$ according to
\begin{equation}\label{arm selection policy}
 \calS(t)=\mathop{\argmax}_{\calS\in\calP(\calK),|\calS|=N}\sum_{k\in\calS}\left(y_k(t-1)+\sqrt{\frac{(N+1)\ln t}{z_k(t-1)}}\right),
\end{equation}
where $y_k(t-1)$ and $\sqrt{(N+1)\ln t/ z_k(t-1)}$ correspond to exploitation and exploration, respectively. Note that the size of $\calP(\calK)$ in exponential in $K$. Thus, the complexity of the combinatorial problem \eqref{arm selection policy} can be very high and some efficient algorithms should be developed \cite{gupta2019multiarmed}. Thanks to the special structure of linear reward, where the rewards of different arms  in \eqref{arm selection policy} are independent on each other, we can find the best super arm  $\calS(t)$ by iteratively choosing the best individual arms. More specifically, we can sort the clients in descending order
of the rewards and choose the top $N$ clients.  In what follows, we provide the analysis of the upper bound of regret $\Sigma^{\pi_1}$ which is logarithmic in $T$ and polynomial in $K$.

\vspace{-0.5cm}
\subsection{Upper Bound on Regret}
\begin{theorem}\label{upper bound theorem 1}
The expected regret achieved by the proposed CS-UCB algorithm presented in \textbf{Algorithm  \ref{alg1}} is upper bounded by
\begin{equation}\label{upper bound 1}
  \Sigma^{\pi_1}<\Delta_{\max}\left[\frac{K}{N}+1+\frac{4N^2(N+1)K\ln T}{(\Delta_{\min})^2}+K+\frac{\pi^2}{3}NK\right],
\end{equation}
where $\Delta_{\max}=\max_{r(\calS)\leq r(\calS^{\ast})} \Delta_{\calS}$,  $\Delta_{\calS}=\sum_{k\in\calS^{\ast}}\mu_k-\sum_{k\in\calS}\mu_k$, and $\Delta_{\min}=\min_{r(\calS)\leq r(\calS^{\ast})} \Delta_{\calS}$.
\end{theorem}

\begin{IEEEproof}
The sketch of the proof is provided in Appendix \ref{proof of upper bound 1} and refer to \cite{gai2012combinatorial} for more details.
\end{IEEEproof}

According to \textbf{Theorem \ref{upper bound theorem 1}}, the regret grows as  $\mathcal{O}(N^3K\ln T)$, i.e., polynomially in the number of wireless channels, linearly in the number of clients, and strictly logarithmically in the number of communication rounds. Given that $N$ and $K$ are fixed, the gap between the optimal solution and the proposed CS-UCB algorithm becomes smaller in  the logarithmic way in $T$.
\vspace{-0.5cm}
\subsection{Convergence Analysis}
In the ideal  scenario, it is assumed that the local datasets of the clients are balanced and the sample points in each $\calW_k$ are sampled i.i.d. from the same source distribution $\mathcal{W}$, i.e., $\mathcal{W}_k\subseteq \mathcal{W}, \forall k\in\mathcal{K}$. Thus, the weights $\varpi_k$'s are set with the same value, i.e., $\varpi_k=\frac{1}{N},\forall k\in \calK$. In the following, we first highlight the necessary assumptions and then provide the convergence guarantee.
\begin{assumption}\label{Unbiased evaluation}
(Unbiased evaluation) Given $g_{\hat{s}}(\qx)=\sum_{s^{\prime}=1}^{\hat{s}}g(\qx;\qw_{s^{\prime}})$, where $\qw_{s^{\prime}}$'s are i.i.d sampled from $\calW$. $g_{\hat{s}}$ is  an unbiased estimator of the global loss function $G$, i.e., $\Ex[g_{\hat{s}}(\qx)]=G(\qx)$.
\end{assumption}
\begin{assumption}\label{L-smoothness}
(L-smoothness) $G(\qx)$ is L-smooth, i.e., $G(\qx^{\prime})-G(\qx)\leq \nabla  G^\dag(\qx)(\qx^{\prime}-\qx)+\frac{L}{2}||\qx^{\prime}-\qx||_2^2$.
\end{assumption}
\begin{assumption}\label{Strong convexity}
(Strong convexity) $G(\qx)$ is $\Phi$-strongly convex, i.e., $G(\qx^{\prime})-G(\qx)\geq\nabla  G(\qx)^\dag(\qx^{\prime}-\qx)+\frac{\Phi}{2}||\qx^{\prime}-\qx||_2^2$.
\end{assumption}
\begin{assumption}\label{Bounded variance}
(Bounded variance \cite{xie2018zeno}) It is assumed that any correct gradient estimator $\nabla G_k(\qx(t))$ in any
round $t$ has upper bounded variance, i.e., $\Ex ||\nabla G_k(\qx(t))-\Ex[\nabla G_k(\qx(t))]||^2_2\leq \delta_0, \forall k\in\calK$.
\end{assumption}
\begin{theorem}\label{convergence theorem 1}
Under \textbf{Assumptions \ref{Unbiased evaluation}-\ref{Bounded variance}} and taking $\gamma\leq\frac{1}{L}\leq\frac{1}{\Phi}$, we have
\begin{equation}
 \Ex[G(\qx(T))-G(\qx^{\ast})]\leq \frac{\delta_0}{2N\Phi} + (1-\gamma\Phi)^{T-1}\bigg[G(\qx(1))]-G(\qx^{\ast})-\frac{\delta_0}{2N\Phi}\bigg],
\end{equation}
where $\qx^{\ast}$ denotes the optimal weights.
\end{theorem}
\begin{IEEEproof}
See Appendix \ref{proof of convergence theorem 1} for reference.
\end{IEEEproof}

It is observed from \textbf{Theorem \ref{convergence theorem 1}} that as $N$ increases, the gap between $G(\qx(T))$ and $G(\qx^{\ast})$ becomes smaller, suggesting a better convergence performance.
\vspace{-0.5cm}
\section{Scheduling in Non-ideal Scenario}\label{section scheduling in NS}
In this section, problem \eqref{original problem}  is investigated which takes both the availability constraint and the fairness constraint into account. Similar to problem \eqref{original pro}, we first reformulate  \eqref{original problem} as an MAB problem, but \textbf{Algorithm \ref{alg1}} cannot be applied to problem \eqref{original problem} directly because of  the fairness constraint. The availability constraint can be addressed by checking the availability of clients before choosing an action in \textbf{Algorithm 1} \cite{auer2002finite}, but satisfying the fairness constraint has to resort to other approaches. Motivated by \cite{neely2010stochastic,li2019combinatorial}, we apply the virtual queue technique to handle the fairness constraint.
\vspace{-0.5cm}
\subsection{Problem Reformulation}
To address the uncertainty of the set of the  available clients, we consider the special class of stationary and randomized policies named $\mathcal{A}$-only policies, which observe $\calA(t)$ in each round $t$ and independently select a super arm $\mathcal{S}(t)\in\mathcal{Q}(\calA(t))$ as a pure (possibly randomized) function of the observed $\mathcal{A}(t)$ only \cite{neely2010stochastic}.  We define a vector of probability distributions $\mathbf{q}=[q_{\calS}(e),\forall \calS\in\calQ(e),\forall e\in\calP(\calK)]$ to describe an $\calA$-only policy $\pi$ where $q_{\calS}(e)$ is the probability that the super arm $\calS$ is chosen when the set $e\in\calP(\calK)$ of the available arms  is observed. Note that $\sum_{\calS\in\calQ(e)}{q_{\calS}(e)}=1, \forall e\in\mathcal{P}(\calK)$. Then, based on the $\calA$-only  policy $\pi$, the mean of $b_k(t)$ for all $t$'s is given as
\begin{equation}
  \Ex[b^{\pi}_k(t)]=\sum_{e\in\calP(\calK)}\hat{P}_{\calA}(e)\sum_{\calS\in\calQ(e):k\in\calS}q_{\calS}(e),
\end{equation}
and the fairness constraint \eqref{fairness constraint} is equivalent to $ \Ex[b^{\pi}_k(t)]\geq c_k$. Furthermore,  the following lemma is obtained.

\begin{lemma}[\cite{neely2010stochastic}]\label{lemma 1}
There always exists an $\calA$-policy $\pi$ that can meet the fairness constraint specified by a vector $\mathbf{c}$, i.e.,
 \begin{equation}
  \Ex[b^{\pi}_k(t)]\geq c_k, \forall k\in\calK,
\end{equation}
if $\mathbf{c}$ is strictly inside the maximal feasibility region $\calC$.
\end{lemma}
\begin{IEEEproof}
The proof is very similar to the proof of Theorem 4.5 in \cite{neely2010stochastic} and hence is omitted here.
\end{IEEEproof}
Assume that the  expectation $\mu_k$ of the normalized reward of each arm $k$ is known in advance,  the normalized  reward $\mu(\calS)$ of the super arm $\calS$ is a constant.    Based on \textbf{Lemma 1}, we can reformulate problem  \eqref{original problem} as a linear program:
\begin{equation}\label{linear problem}
\begin{split}
  \mathop{\max}_{\qq} \ & \sum_{e\in\calP(\calK)}\hat{P}_{\calA}(e)\sum_{\calS\in\calQ(e)}q_{\calS}(e)\mu(\calS)\\
  \text{s.t.} \ &\sum_{e\in\calP(\calK)}\hat{P}_{\calA}(e)\sum_{\calS\in\calQ(e):k\in\calS}q_{\calS}(e)\geq c_k,\forall k\in\calK,\\
  &\sum_{\calS\in\calQ(e)}{q_{\calS}(e)}=1, \forall e\in\calP(\calK),\\
  &q_{\calS}(e)\in[0,1], \forall \calS\in\calQ(e), \forall e\in\calP(\calK),
\end{split}
\end{equation}
whose optimal solution is easy to  find  due to its linearity. However, the linear program is based on the assumption that the normalized expectation of each arm's reward is known in advance, which is usually impractical. Thus, to achieve the maximal reward, the player (i.e., the AP) needs to estimate the mean rewards of the arms and exploit such knowledge to find out a super arm that seems to yield the highest reward, but also keep exploring further the other arms to identify with better precision which super arm is actually the best. Such a learning process is a typical exploration-exploitation tradeoff and inevitably leads to the reward loss, i.e., the regret \cite{lai1985asymptotically}.
Similar to problem \eqref{reward mini1}, we use the regret as the metric to evaluate the performance.  Given an optimal $\calA$-policy $\pi^\ast$, we define $r^{\ast}$ as the maximal reward of problem \eqref{linear problem} with  known reward expectation $\mu_k,k\in\calK$, which is given as
\begin{equation}
 r^{\ast}=\sum_{e\in\calP(\calK)}\hat{P}_{\calA}(e)\sum_{\calS\in\calQ(e)}q^{\ast}_{\calS}(e)\mu(\calS),
\end{equation}
where $q^{\ast}_{\calS}$ is the associated probability distribution of policy $\pi^\ast$. Assume that the FL training converges at the end of round $T$, i.e., $\eta(T)=1$ and $\eta(T+1)=0$.  Then, problem  \eqref{original problem} is equivalent to the minimization of the cumulative regret under a policy $\pi_2$ by choosing a super arm $\calS(t)$ in each round $t$, i.e.,
\begin{equation}\label{regret problem}
\begin{split}
\min_{\{\calS(t),t\geq 1\}} \  \Sigma^{\pi_2}=   Tr^{\ast}-\Ex\bigg[\sum_{t=1}^{T} r(\calS(t))\bigg], \ \ \text{s.t.} \ \eqref{fairness constraint} \ \text{and} \ \eqref{availability constraint3}.
\end{split}
\end{equation}
\vspace{-0.1cm}
\subsection{Proposed Algorithm}
\begin{algorithm}[htb]\small
\caption{Proposed CS-UCB-Q algorithm for problem  \eqref{regret problem}.}
\label{alg2}
\begin{algorithmic}[1]
\STATE \textbf{Initialization:} Set $z_k(1)=0$ and $D_k(1)=0, \forall k\in\calK$.
\STATE \textbf{Main loop:}
\FOR{$t=1,\ldots,T$}
\FOR{$k\in\calK$}
\STATE \textbf{if} $z_k(t)>0$ \textbf{then} update $\hat{y}_k(t)$ according to \eqref{truncated ucb},
\STATE \textbf{else} set $\hat{y}_k(t)=1$. \textbf{end if}
\STATE Update $D_k$ according to \eqref{queue}.
\ENDFOR
\STATE Choose a super arm $\calS(t)$ from $\calA(t)$ according to \eqref{arm selection} and update $b_k(t), \forall k\in\calK$.
\STATE Update $\mathbf{y}(t)$ and $\mathbf{z}(t)$ according to  \eqref{update y} and \eqref{update z}, respectively.
\ENDFOR
\end{algorithmic}
\end{algorithm}

As shown in \textbf{Algorithm \ref{alg2}}, the proposed algorithm for problem \eqref{regret problem} is based on policy UCB \cite{auer2002finite}  and virtual queue technique \cite{neely2010stochastic}, which is referred to as CS-UCB-Q algorithm.  Recall that $y_k(t)$ is the the empirical average of the observed reward of arm $k$ up to the current round $t$ and $z_k(t)=\sum_{t^{\prime}=1}^{t}b_k(t^{\prime})$ is the number of times that arm $k$ has been played up to the current round $t$.  Then, we use a truncated UCB estimate to calculate the reward of each arm $k$,
\begin{equation}\label{truncated ucb}
  \hat{y}_k(t)=\min\{y_k(t-1)+\sqrt{\frac{2\ln t}{z_k(t-1)}},1\}.
\end{equation}
Note that $\hat{y}_k(t)=1$, if $z_k(t-1)=0$.

To deal with the fairness constraint, we introduce a virtual queue $D_k$ for each arm $k$ as follows
\begin{equation}\label{queue}
 D_k(t)=[D_k(t-1)+c_k-b_k(t-1)]^{+},
\end{equation}
where $[x]^{+}=\max\{x,0\}$ and $ D_k(t)$ indicates the queue length at the beginning of round $t$. We define $\qD(t)=[D_1(t), D_2(t),\ldots,D_K(t)]$ and initialize $\qD(1)=\mathbf{0}$.  The queue length  decreases by 1 if arm $k$ is selected in round $t-1$, i.e., $b_k(t-1)=1$, and  increases by $c_k$ in each round.  Then, we give the criterion of choosing a super arm $\calS(t)$ in each round $t$ which maximizes the compound value of $ \hat{y}_k(t)$ and $D_k(t)$, i.e.,
\begin{equation}\label{arm selection}
  \calS(t) \in \argmax_{\substack{\calS\in\calQ(\calA(t)),\\ |\calS|=\min\{N, |\calA(t)|\}}}\sum_{k\in\calS} (1-\beta)\hat{y}_k(t)+\beta D_k(t),
\end{equation}
where the weighting factor $\beta\in[0,1]$ is a non-negative  constant.  Similar to \eqref{arm selection policy}, the compound rewards of different clients are independent on each other, we can select the best individual arms iteratively.
\vspace{-0.5cm}
\subsection{Upper Bound on Regret}
In this subsection, we first validate that  the proposed CS-UCB-Q algorithm is feasibility-optimal. Then we provide an upper bound  on the expected regret of the proposed CS-UCB-Q algorithm.

\begin{theorem}\label{feasibility optimality theorem}
The proposed CS-UCB-Q algorithm presented in \textbf{Algorithm  \ref{alg2}} is feasibility-optimal. In other words, the fairness constraint in \eqref{fairness constraint} is satisfied, for any $\qc$ strictly in the maximal feasibility region $\calC$.
\end{theorem}
\begin{IEEEproof}
See Appendix \ref{proof of feasibility optimality theorem} for reference.
\end{IEEEproof}
According to  \textbf{Theorem \ref{feasibility optimality theorem}}, the long-term fairness constraint can be satisfied under the proposed CS-UCB-Q algorithm presented in \textbf{Algorithm  \ref{alg2}} as long as the
requirement is feasible, i.e., $\qc$   is strictly in the maximal feasibility region $\calC$.

\begin{theorem}\label{upper bound theorem 2}
The expected regret achieved by the proposed CS-UCB-Q algorithm presented in \textbf{Algorithm  \ref{alg2}} is upper bounded by
\begin{equation}\label{upper bound2}
 \Sigma^{\pi_2}\leq  \beta\frac{\Omega T}{2}+(1-\beta)\Big[(\frac{\pi^2}{3}+1)K+4\sqrt{2KNT\ln T}\Big],
\end{equation}
where $\Omega=\sum_{k\in\calK}\max\{c_k^2,(1-c_k)^2\}$. Typically when $0<\beta\leq\frac{1}{\sqrt{T}}$ and $T$ is large enough, we have
\begin{equation}
 \Sigma^{\pi_2}\leq  \frac{\Omega \sqrt{T}}{2}+(\frac{\pi^2}{3}+1)K+4\sqrt{2KNT\ln T}.
\end{equation}
\end{theorem}
\begin{IEEEproof}
The sketch of the proof is provided in Appendix \ref{proof of upper bound theorem 2} and refer to \cite{hsu2018intergrating} for more details.
\end{IEEEproof}

Note that the regret bound in \eqref{upper bound2} has two terms where $\beta$ is attributed to achieve a balance between them. Specifically, when $\beta$ is large, i.e., a higher priority to satisfying the fairness constraint, the regret bound mainly depends  on the first term and approximately is  of the order  $\mathcal{O}{(T)}$. However, a small $\beta$ (e.g., $\beta\leq\frac{1}{\sqrt{T}},T\rightarrow +\infty$) suggests that the regret bound mainly depends  on the second term and approximately is  of the order  $\mathcal{O}{(\sqrt{T \ln T})}$. In this case, the performance gap between the optimal solution and the solution found  by the proposed CS-UCB-Q becomes smaller in the sub-linear way over communication rounds.
\vspace{-0.5cm}
\subsection{Convergence Analysis}
Different from the ideal scenario, the datasets of the clients in the  non-ideal scenario are heterogeneous.
Here, we adopt a heuristic way to set the weights as $\varpi_k(t)=\frac{s_k}{\sum_{k^{\prime}\in\calK}s_{k^{\prime}}}$ \cite{mcmahan2016communication}. To prove the convergence, we need another assumption as follows:
\begin{assumption}\label{The first and second moment conditions}
(The first and second moment conditions \cite{chen2020efficient}) The global loss function $G(\qx)$ and the
aggregation operation $\qv(t)$ satisfy the following:
\begin{description}
  \item[(a)] The global loss function $G(\qx)$ is bounded by a scalar $G^{\inf}=G(\qx^{\ast})$.
  \item[(b)] Given two scalars $\delta_1>\delta_2>0$, two constraints $||\Ex[\qv(t)]||_2\leq\delta_1||\nabla G(\qx(t))||_2$ and $\nabla G^\dag(\qx(t))\Ex[\qv(t)]\geq \delta_2|| G(\qx(t))||_2^2$ always hold.
  \item[(c)] There exists $\delta_3$ such that $\Ex||\qv(t)||_2^2-||\Ex[\qv(t)]||_2^2\leq\delta_3+||\nabla G(\qx(t))||_2^2$.
\end{description}
\end{assumption}
\begin{theorem}\label{convergence theorem 2}
Under \textbf{Assumptions \ref{L-smoothness}, \ref{Strong convexity}}, and \textbf{\ref{The first and second moment conditions}} and taking a fixed learning rate satisfying $\gamma\leq\frac{\delta_2}{L(1+\delta_1^2)}$ with $L \geq \Phi$, we have
\begin{align}
\Ex[G(\qx(T))-G(\qx^{\ast})]\leq \frac{\gamma\delta_3L}{2\Phi\delta_2}+(1-\gamma\delta_2\Phi)^{T-1}\Ex\bigg[G(\qx(1))]-G(\qx^{\ast})-\frac{\gamma\delta_3L}{2\Phi\delta_2}\bigg].
\end{align}
\end{theorem}
\begin{IEEEproof}
See Appendix \ref{proof of convergence theorem 2} for reference.
\end{IEEEproof}

According to \textbf{Theorem 5}, a smaller $\delta_3$ value leads to a  better convergence performance. To decrease $\delta_3$ value, we can increase the number $|\calS(t)|$ of the participating clients in each round if possible. But their relationship is not so tight and thus the convergence rate has a weak dependence on $|\calS(t)|$. Besides,  we can involve the clients with important local datasets  in more communication rounds to reduce $\delta_3$ value, which is realized by using the fairness constraint.

\vspace{-0.3cm}
\section{Numerical Results}\label{section numerical results}
In this section, we conduct experiments to evaluate the performance of the proposed CS-UCB and CS-UCB-Q algorithms. As mentioned in Section \ref{section system model}, the time cost $\tau_{k}(t)$ of each client $k$ in round $t$ is directly observed by the AP in practice. In our simulations,  in order to obtain the value of $\tau_{k}(t)$, we first give the expressions to compute the three main components $\tau^{\textrm{D}}_k(t)$, $\tau^{\textrm{LU}}_k(t)$,  and  $\tau^{\textrm{U}}_k(t)$, respectively.

The average distribution time $\tau^{\textrm{D}}_k(t)$ of client $k$ in round $t$ is computed as
\begin{equation}
 \tau^{\textrm{D}}_k(t)=\frac{m_0}{BR^{\textrm{D}}_k(t)},
\end{equation}
where $m_0$ (in bits) is the size of  data for distribution including the global model parameters, $B$ (in Hz) is the bandwidth of one channel, and $R^{\textrm{D}}_k(t)$ (in bits/s/Hz) is the average distribution rate for client $k$ in round $t$. It is worth noting that the coherence time of  wireless channels is relatively short, thus  the distribution process  may experience multiple coherence time slots in a communication round.  In other words, the wireless channel varies during the distribution process. Here,  we use the average distribution rate in a communication round instead of the instantaneous one because we  focus on accelerating the learning process from the long-term learning perspective.  The average distribution rate $R^{\textrm{D}}_k(t)$ in round $t$ is expressed as
\begin{equation}\label{distribution rate}
  R^{\textrm{D}}_k(t)=\log_2\Big[1+\frac{p^{\textrm{D}}_k|h^{\textrm{D}}_k(t)|^2}{\sigma^2}\Big],
\end{equation}
where $p^{\textrm{D}}_k$  is the transmit power of the AP for client $k$, $\sigma^2$ is the  variance of additive  white Gaussian noise, and $h^{\textrm{D}}_k(t)$ is the average downlink channel gain of client $k$ in round $t$.

Similarly, the upload time of client $k$ in round $t$ is
\begin{equation}
 \tau^{\textrm{U}}_k(t)=\frac{m_k}{BR^{\textrm{U}}_k(t)},
\end{equation}
where $m_k$ is the data size of client $k$ for upload and $R^{\textrm{U}}_k(t)$ in bits/s/Hz denotes the average upload rate of client $k$ in round $t$, which is given as
\begin{equation}\label{upload rate}
  R^{\textrm{U}}_k(t)=\log_2\Big[1+\frac{p^{\textrm{U}}_k|h^{\textrm{U}}_k(t)|^2}{\sigma^2}\Big],
\end{equation}
where $p^{\textrm{U}}_k$ is the transmit power of  client $k$  and $h^{\textrm{U}}_k(t)$ is the average uplink channel gain of client $k$ in round $t$.

In addition, the local update time of client $k$ in  round $t$ is calculated as
\begin{equation}
  \tau^{\textrm{LU}}_k(t)=\frac{\varrho_k}{\phi_k(t)}.
\end{equation}
where $\varrho_k$ is the number of sample points for the local updates in each round. For example,  $\varrho_k=s_k$ in the gradient descent method and  $\varrho_k=\varrho,\forall k\in\calK,$ in the stochastic gradient descent (SGD) method with $\varrho$ being the mini-batch size.
Finally, we can calculate $\tau_{k}(t)$ according to \eqref{round_time_client}.

\vspace{-0.5cm}
\subsection{Experiment Settings}
The system setting  is summarized as follows unless otherwise specified. We consider  a network where a single-antenna AP is located in the centre of the network  with a disc of 500m  and the clients randomly distributed within the coverage of the AP. The channel gains of both the uplink and downlink links are composed of both small-scale fading  and large-scale fading, where the small-scale fading is set as  Rayleigh distribution with uniform variance  and the large scale fading are generated using the path-loss model: $\text{PL}\text{[dB]}=128.1 + 37.6\log_{10}(d)$ with $d$ representing the distance in km. The noise power $\sigma^2$ is -107 dBm and the bandwidth of each orthogonal channel is 15 KHz. Both the uplink and downlink transmit power is 23 dBm, i.e., $p^{\text{U}}_k=p^{\text{D}}_k=23 \ \text{dBm}, \forall k\in\calK$. The computing capability $\phi_k(t)$ of each client $k$ is uniformly distributed in $[\phi^{\text{LB}}_k, \phi^{\text{UB}}_k]$, where $\phi^{\text{LB}}_k$ and $\phi^{\text{UB}}_k$ are the lower bound and upper bound, respectively. Here, we assume $\phi^{\text{LB}}_k=(0.5k+0.5)\times20$ and $\phi^{\text{UB}}_k=(0.5k+1.5)\times20, \forall k\in\calK$. In addition, for simplicity, we assume $m_0=m_k=5\times10^3$ bits and the maximal interval $\tau_{\max}=5$  seconds.

We study image classification of handwritten digits 0-9 in the well-known MNIST dataset using multinomial logistic regression. The training and test sets consist of 60,000 and 10,000 samples, respectively. The SGD method is adopted, where the mini-batch size $\varrho$ is 2 and the learning rate is 0.001.

\subsection{Performance in Ideal Scenario}\label{performance in ideal scenarios}
In the ideal scenario, it is assumed that there are $K=20$ clients and $N=5$  channels available in each communication round. The dataset is shuffled and then partitioned into the clients equally.  Before giving the numerical results,  we introduce two baseline algorithms for comparison \cite{yang2019scheduling}. The first one is a random scheduling algorithm  where the AP uniformly selects  $N$  clients at random in each communication round for model updates. The second one is the round robin algorithm in which the clients are divided into groups each with $N$ clients and the AP consecutively assigns each group to access to the radio channels and update their parameters per communication round. We define $\Theta=\tau_{\max}\Sigma^{\pi_1}$ and $\theta=\Theta/T$ as the cumulative and  average performance gaps, respectively, where $\pi_1$ can be the proposed CS-UCB algorithm or the introduced baseline algorithms. Note that some clients selected in a certain round may  not be able to send to its results to the AP for aggregation, e.g.,  due to the large propagation  delay or  processing delay. According to \eqref{round_time_client}, we refer to client $k$ with $\tau_k=\tau_{\max}=5$ seconds as a failed client and count the cumulative number of the failed clients.

\begin{figure}
  \centering
  \subfigure[]{
    \label{cul_gap_is} 
    \includegraphics[width=0.3\textwidth]{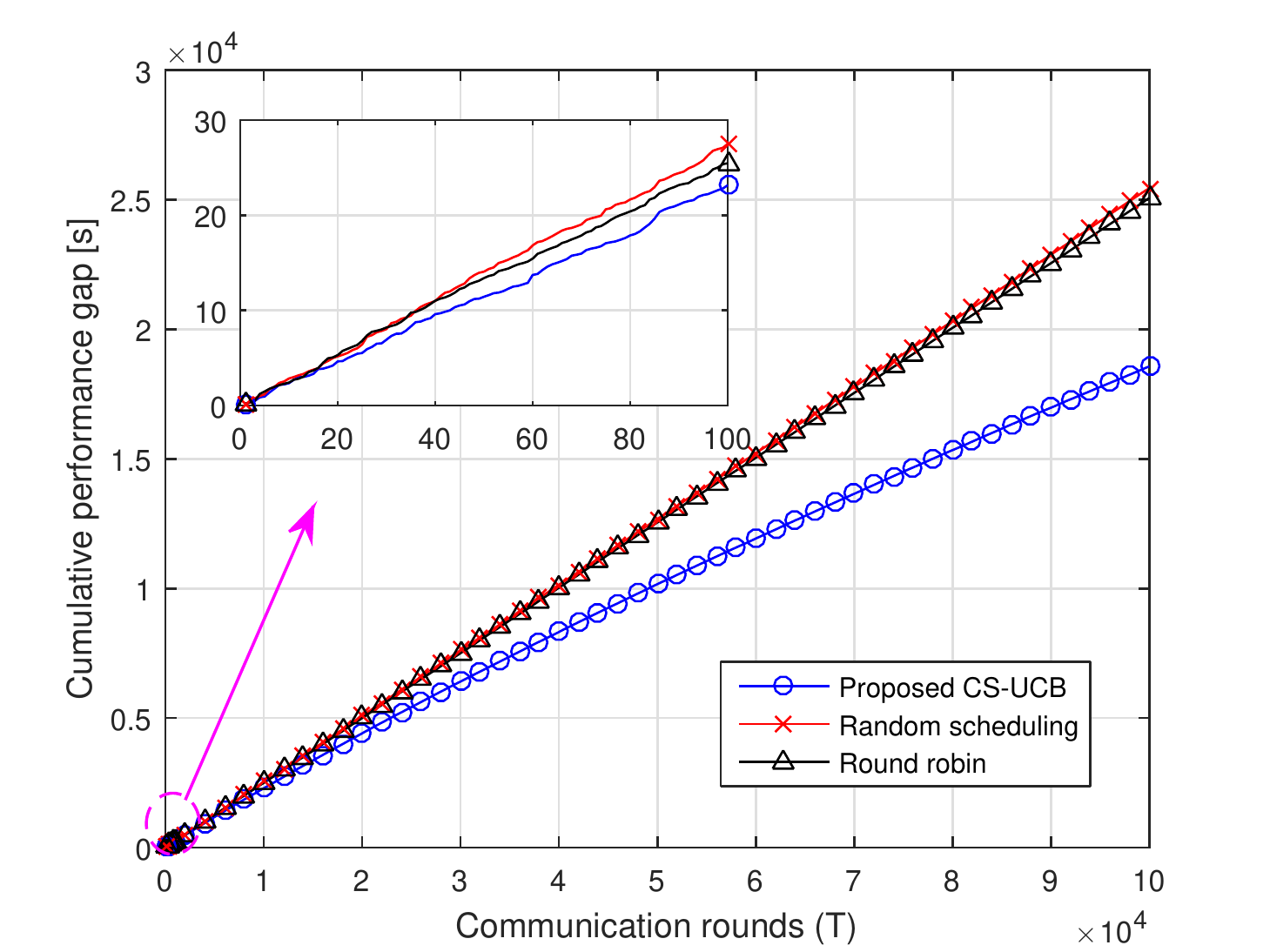}}
  \subfigure[]{
    \label{mean_gap_is} 
    \includegraphics[width=0.3\textwidth]{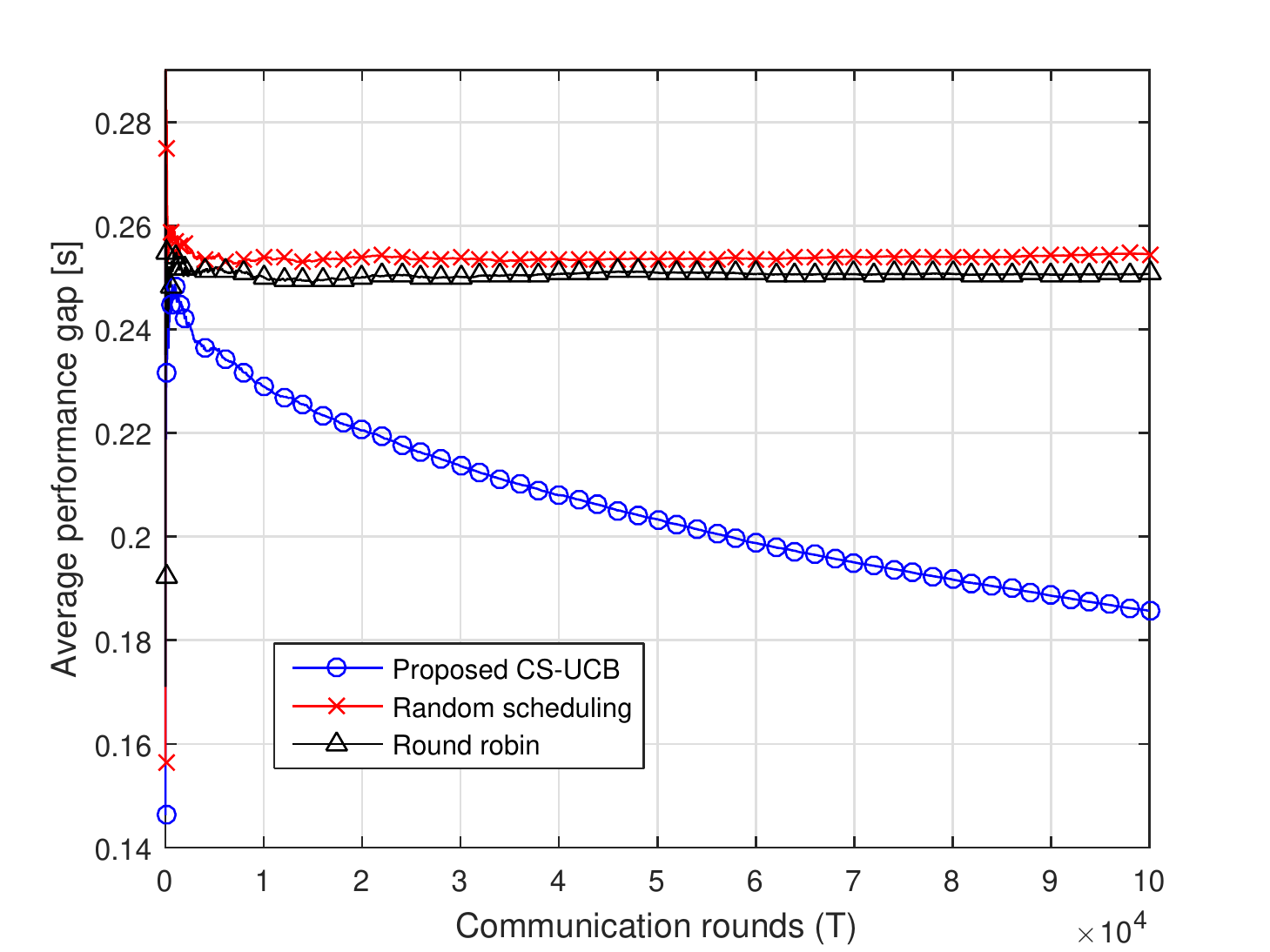}}
      \subfigure[]{
    \label{fail_ratio_is} 
    \includegraphics[width=0.3\textwidth]{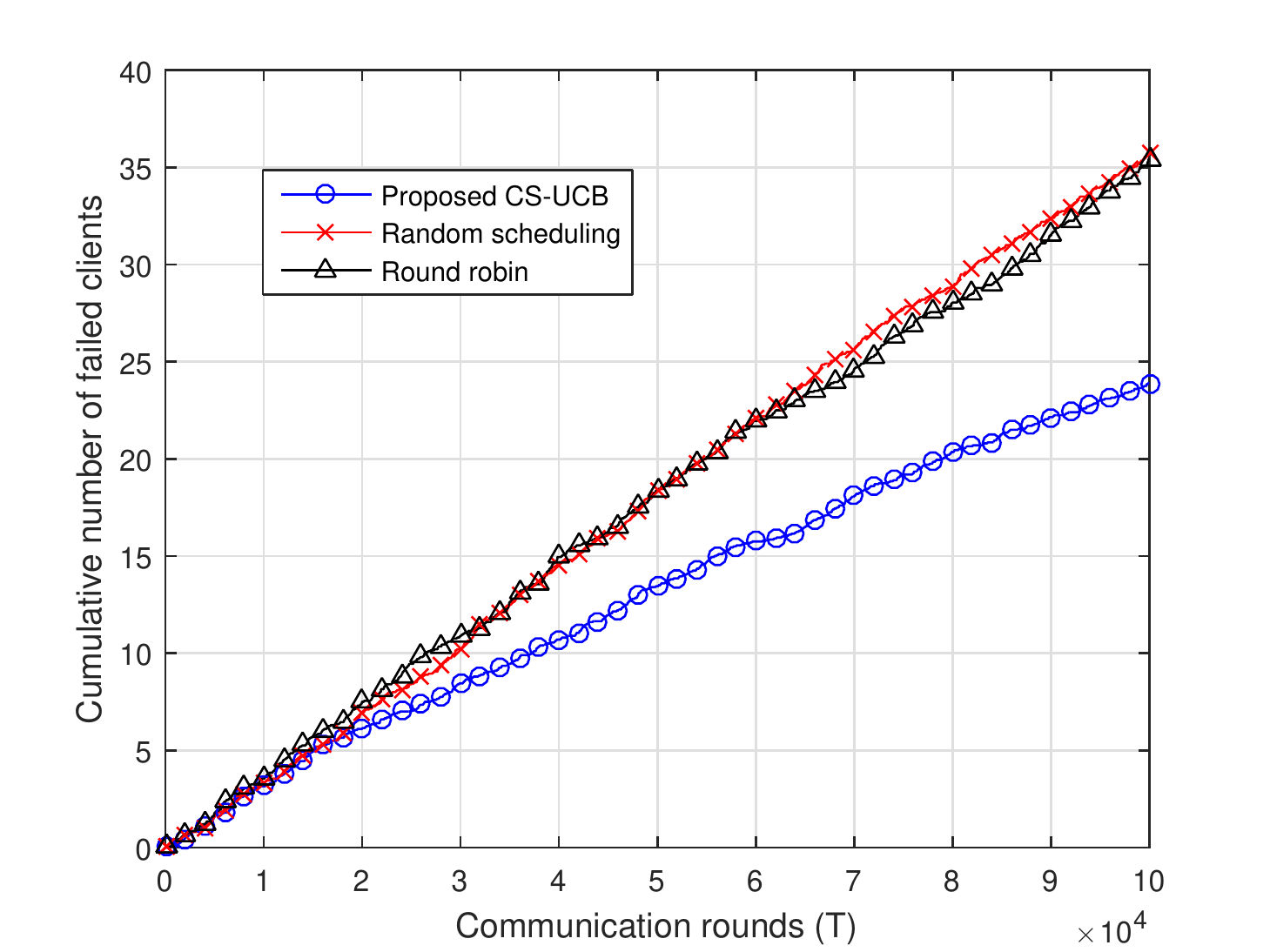}}
  \caption{Performance comparison of different algorithms versus the number of communication rounds: (a) cumulative performance gap, (b) average performance gap, and (c) cumulative number of failed clients with \{$K=20$, $N=5$\}.}
  \label{comparison_round_is}
\end{figure}

As seen from Fig. \ref{comparison_round_is}, the proposed CS-UCB algorithm achieves better performance than the two baseline algorithms after 20 rounds and the performance gap increases as the number of communication rounds increases. This is because the proposed CS-UCB algorithm keeps learning the statistical information of each client in the execution process and leverage a tradeoff between the exploitation of learned knowledge and  the exploration of more potential actions. It is also observed from Fig. \ref{fail_ratio_is}, the proposed CS-UCB algorithm has less  failed clients, which suggests that in a given time interval, more clients can get involved in the learning process for more communication rounds.

\begin{figure}
  \begin{minipage}[t]{0.47\textwidth}
    \centering
    \includegraphics[width=1\textwidth]{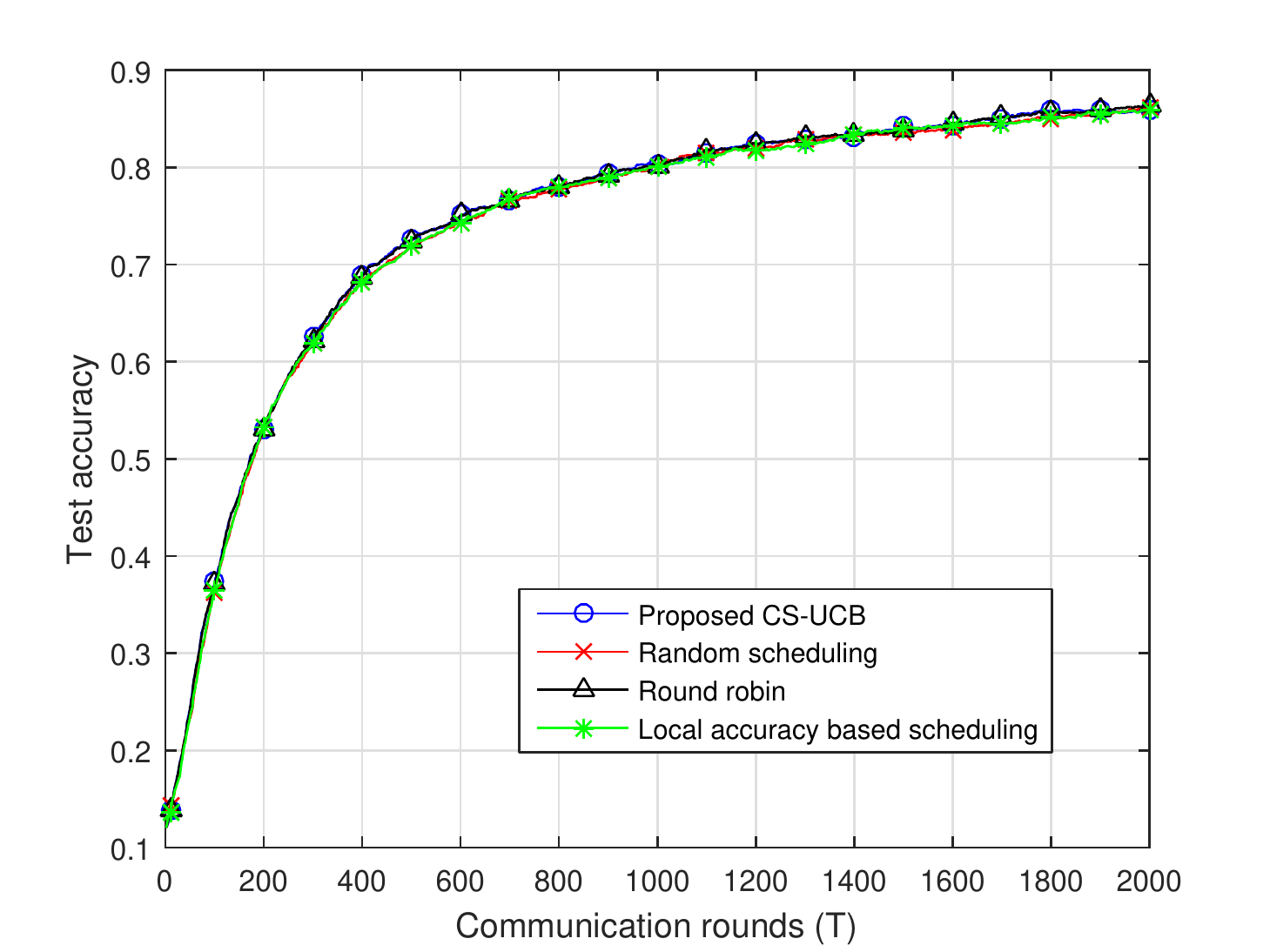}
    \caption{Test accuracy  versus the number of communication rounds of the proposed CS-UCB algorithm  and baselines with \{$K=20,N=5$\}.}
    \label{test_acc_ideal}
  \end{minipage}%
  \hfill
  \begin{minipage}[t]{0.47\textwidth}
    \centering
    \includegraphics[width=1\textwidth]{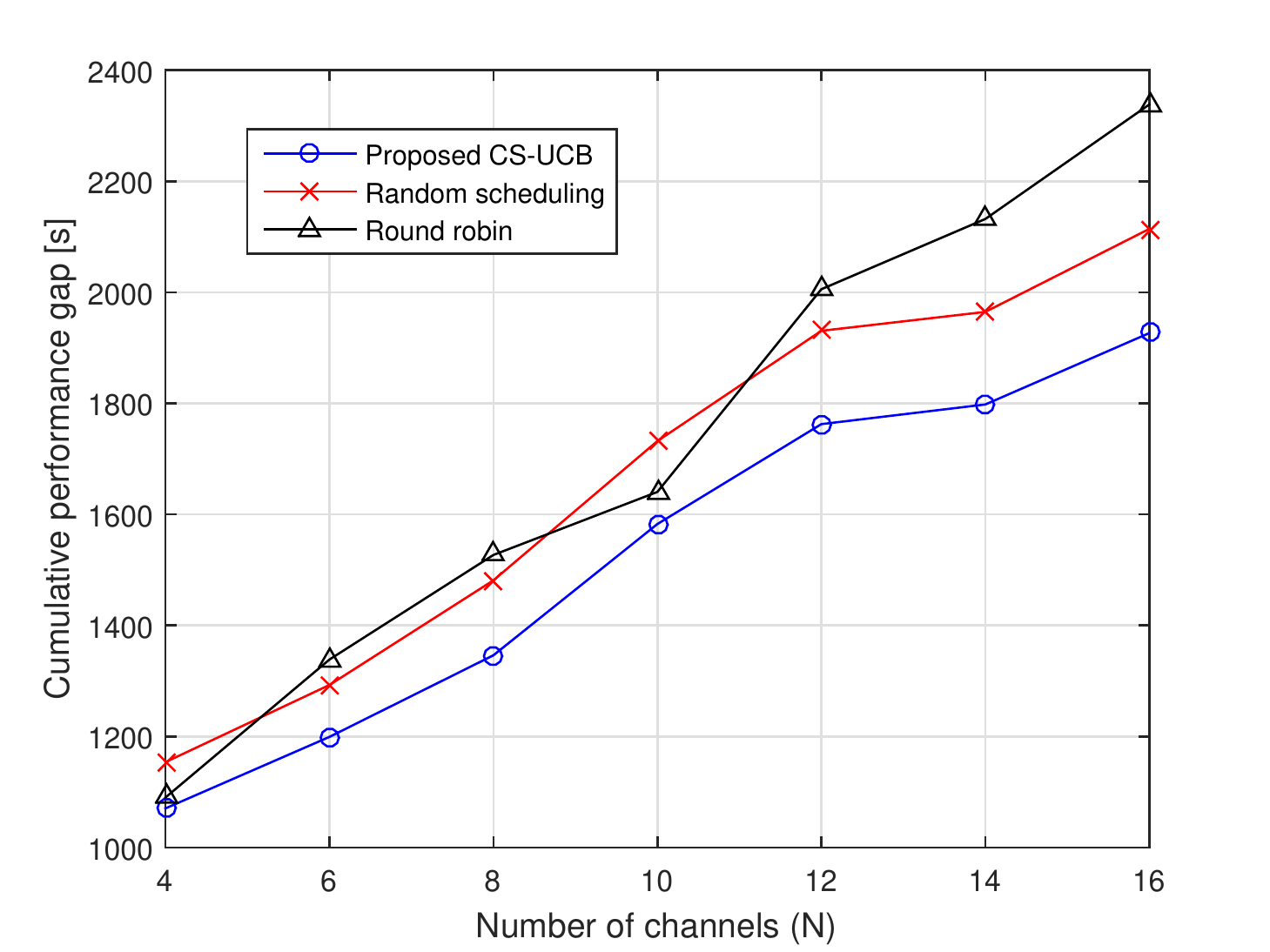}
    \caption{Performance comparison of different algorithms with respect to the number of channels.}
    \label{cum_gap_n_is}
  \end{minipage}
\end{figure}

Before showing the convergence performance, in addition to the existing baseline algorithms, we introduce another baseline algorithm, named local accuracy based scheduling, where the clients with lower local test accuracy have higher priority to be chosen in each round. The test accuracy with respect to  the number of communication rounds of different algorithms is shown in Fig. \ref{test_acc_ideal}. We find that the convergence rate, in terms of the communication rounds, of  the proposed CS-UCB algorithm in the ideal scenario is close to those of the baselines. This is because  the local datasets of the clients are i.i.d. and the average number of the participating clients  in each round is almost the same. Based on the results of  Figs. \ref{comparison_round_is}(a) and \ref{test_acc_ideal}, we can conclude that the proposed  CS-UCB algorithm consumes less wall-clock time to finish the required communication rounds of a certain level of test accuracy.

In Fig. \ref{cum_gap_n_is}, we fix $T=5000$ and show the performance of different algorithms with respect to the number of available channels $N$. It is observed that the performance of the proposed CS-UCB algorithm  is always better than those of the random scheduling algorithm and the round robin algorithm. Besides, we observe that the cumulative performance gap increases as the number of available channels increases because more clients can get involved in the FL process and then the probability of choosing the high-latency clients is increased.

%

\subsection{Performance in Non-Ideal Scenario}
\begin{figure}
  \centering
  \subfigure[]{
    \label{cul_gap_ns} 
    \includegraphics[width=0.3\textwidth]{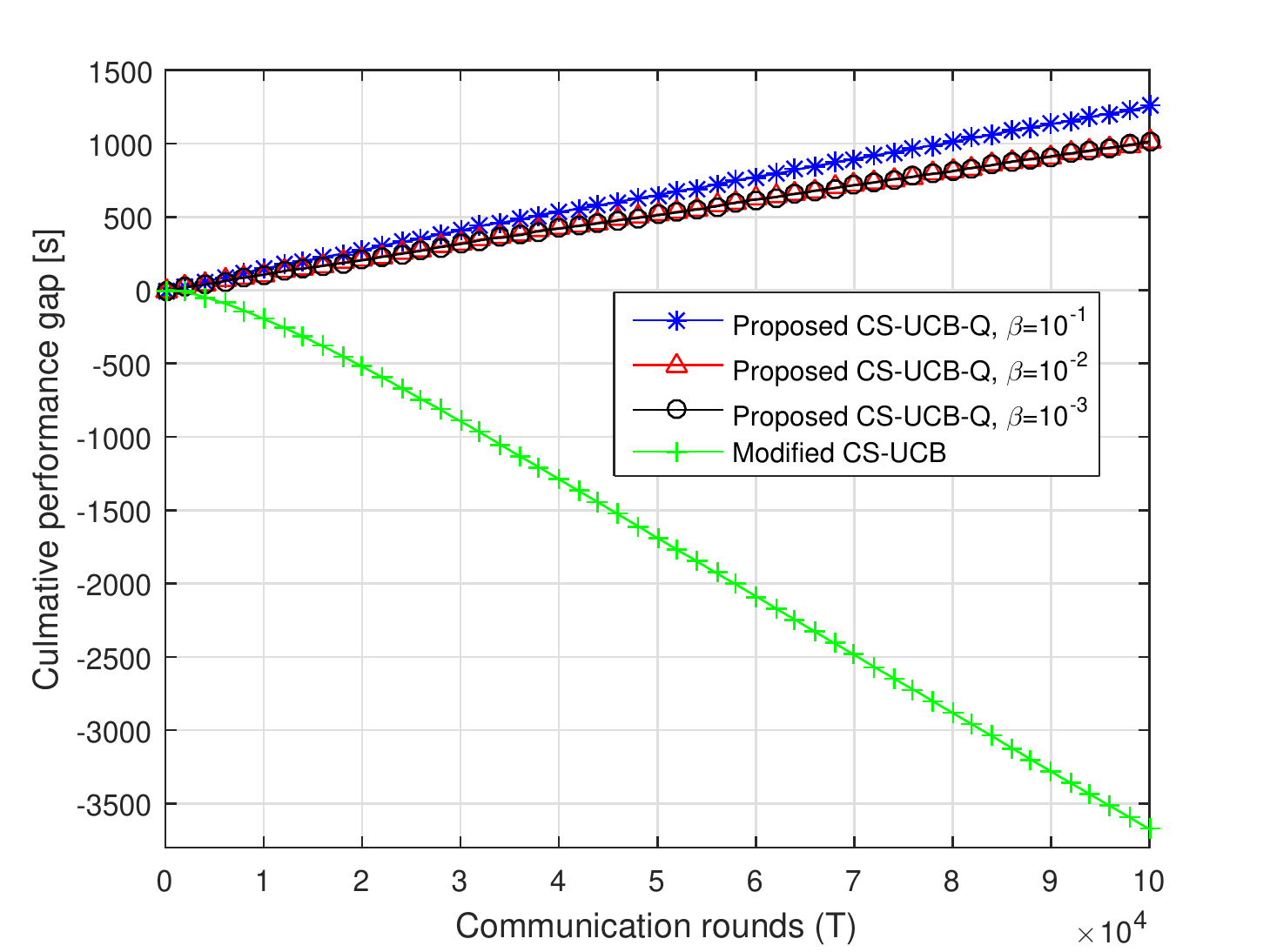}}
  \subfigure[]{
    \label{mean_gap_ns} 
    \includegraphics[width=0.3\textwidth]{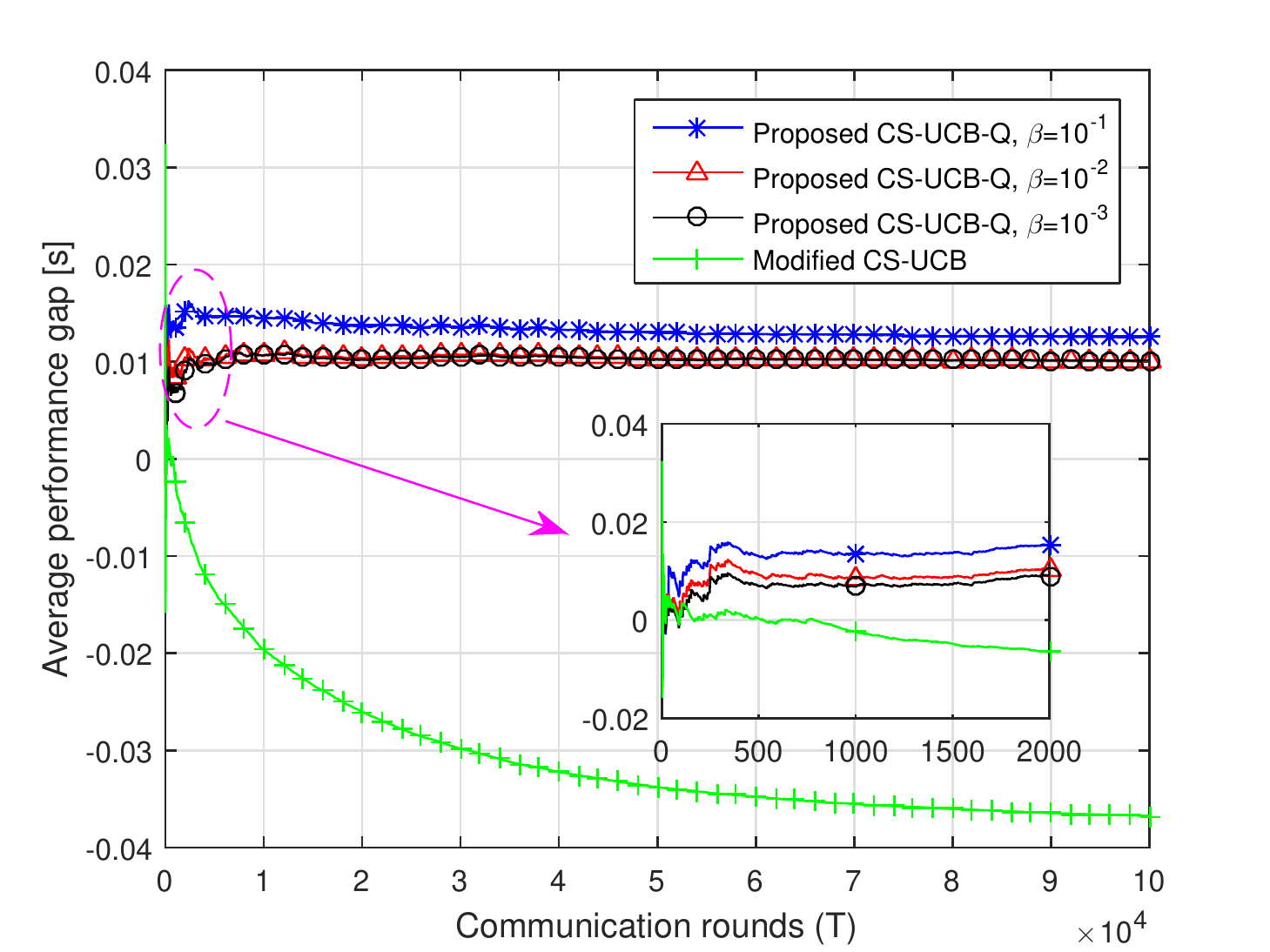}}
      \subfigure[]{
    \label{arm_ratio_bar_nS} 
    \includegraphics[width=0.3\textwidth]{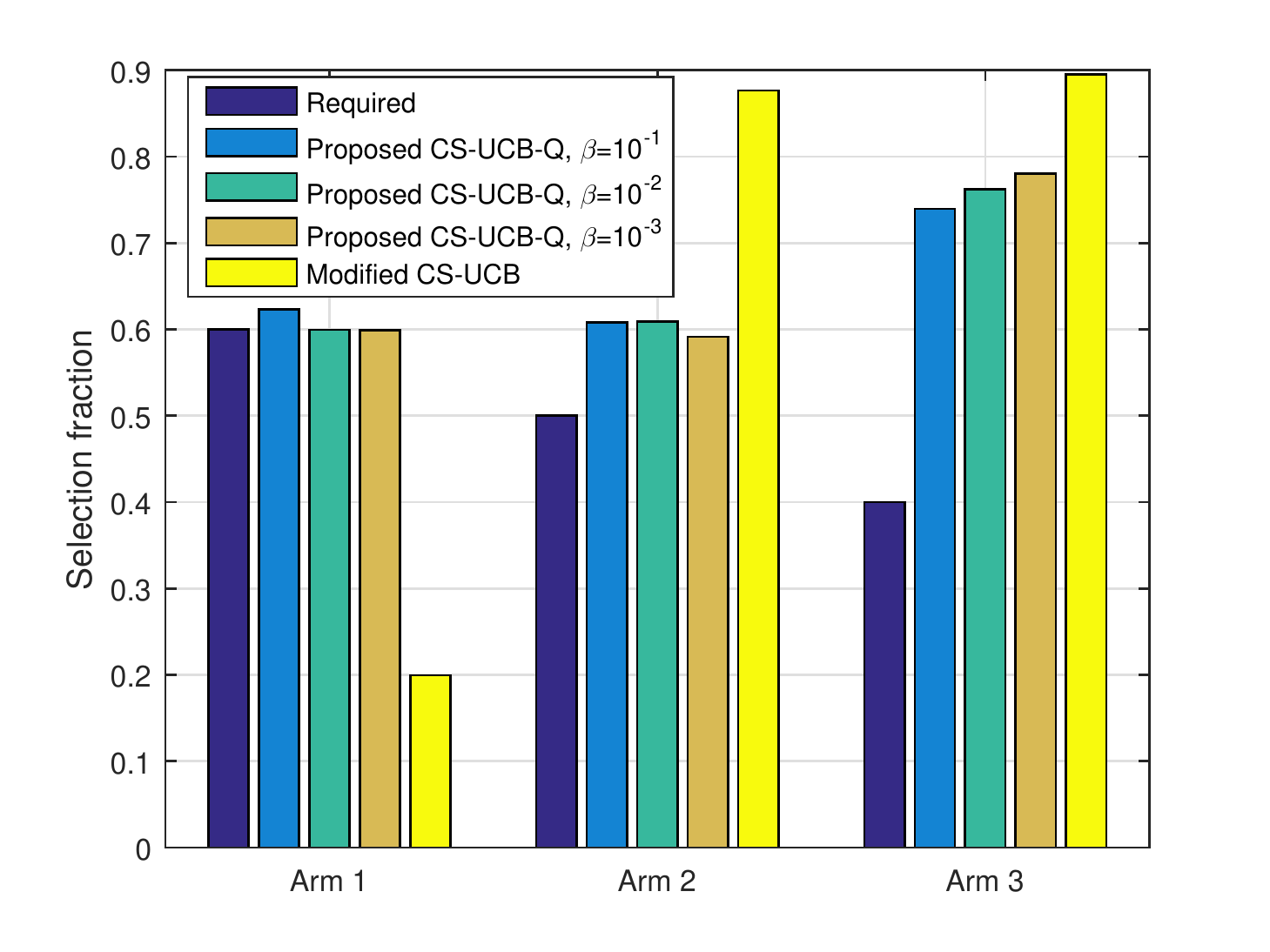}}
  \caption{Performance comparison of different algorithms versus the number of communication rounds: (a) cumulative performance gap, (b) average performance gap, and (c) selection fraction of each client with \{$K=3$, $N=2$\}.}
  \label{comparison_round_ns}
\end{figure}

\begin{figure}
  \centering
  \subfigure[Client 1]{
    \label{arm1} 
    \includegraphics[width=0.3\textwidth]{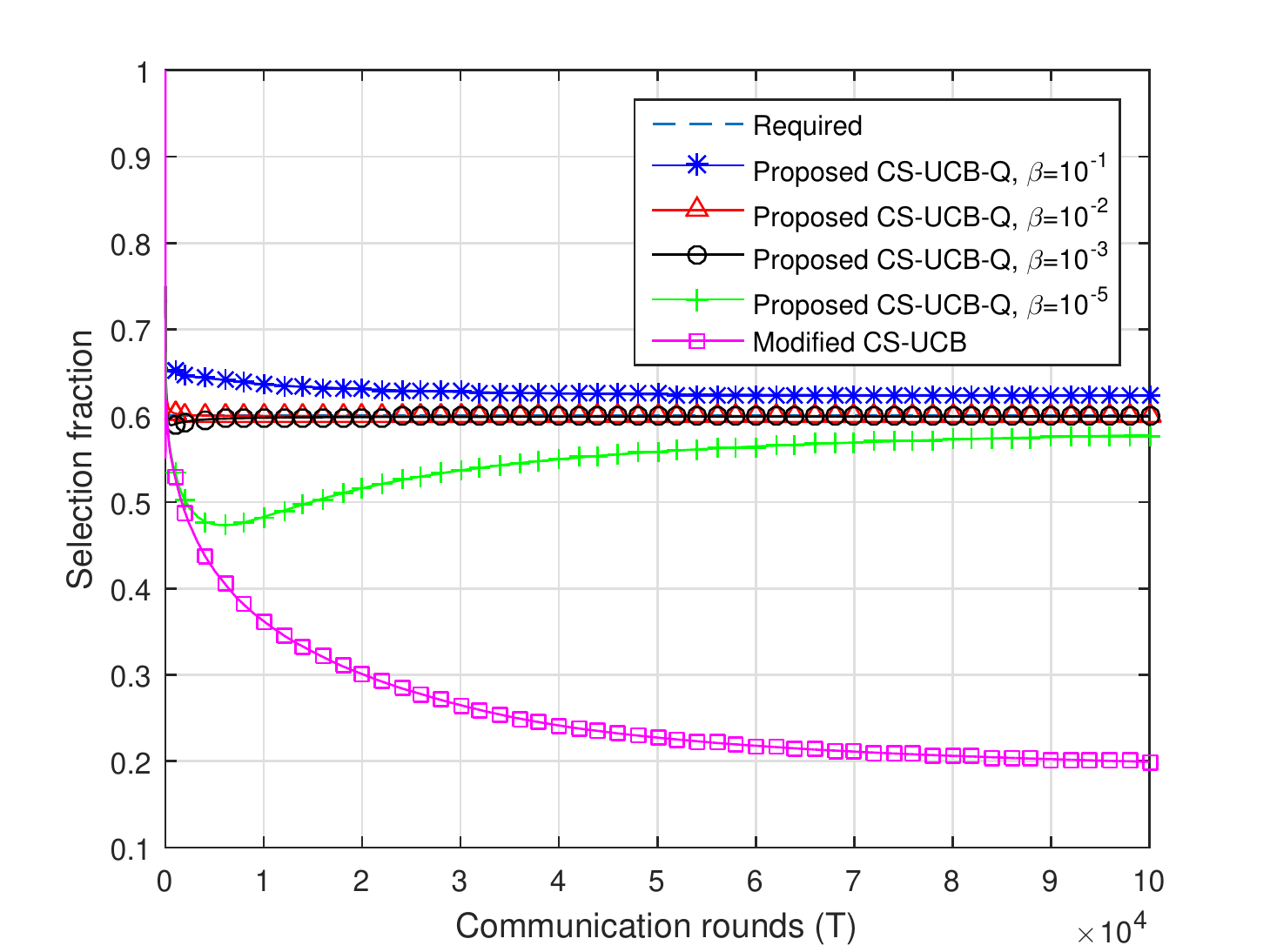}}
  \subfigure[Client 2]{
    \label{arm2} 
    \includegraphics[width=0.3\textwidth]{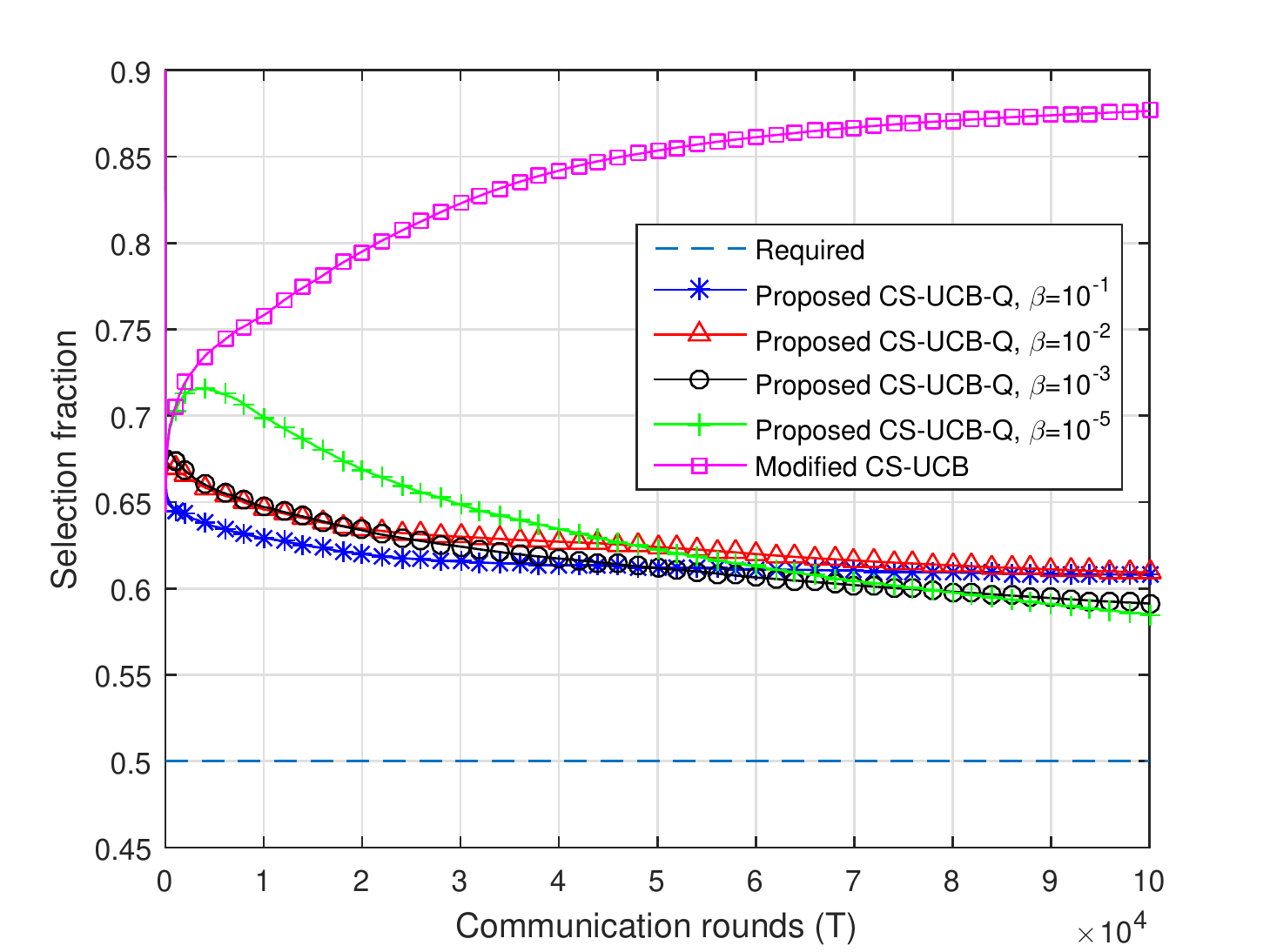}}
      \subfigure[Client 3]{
    \label{arm3} 
    \includegraphics[width=0.3\textwidth]{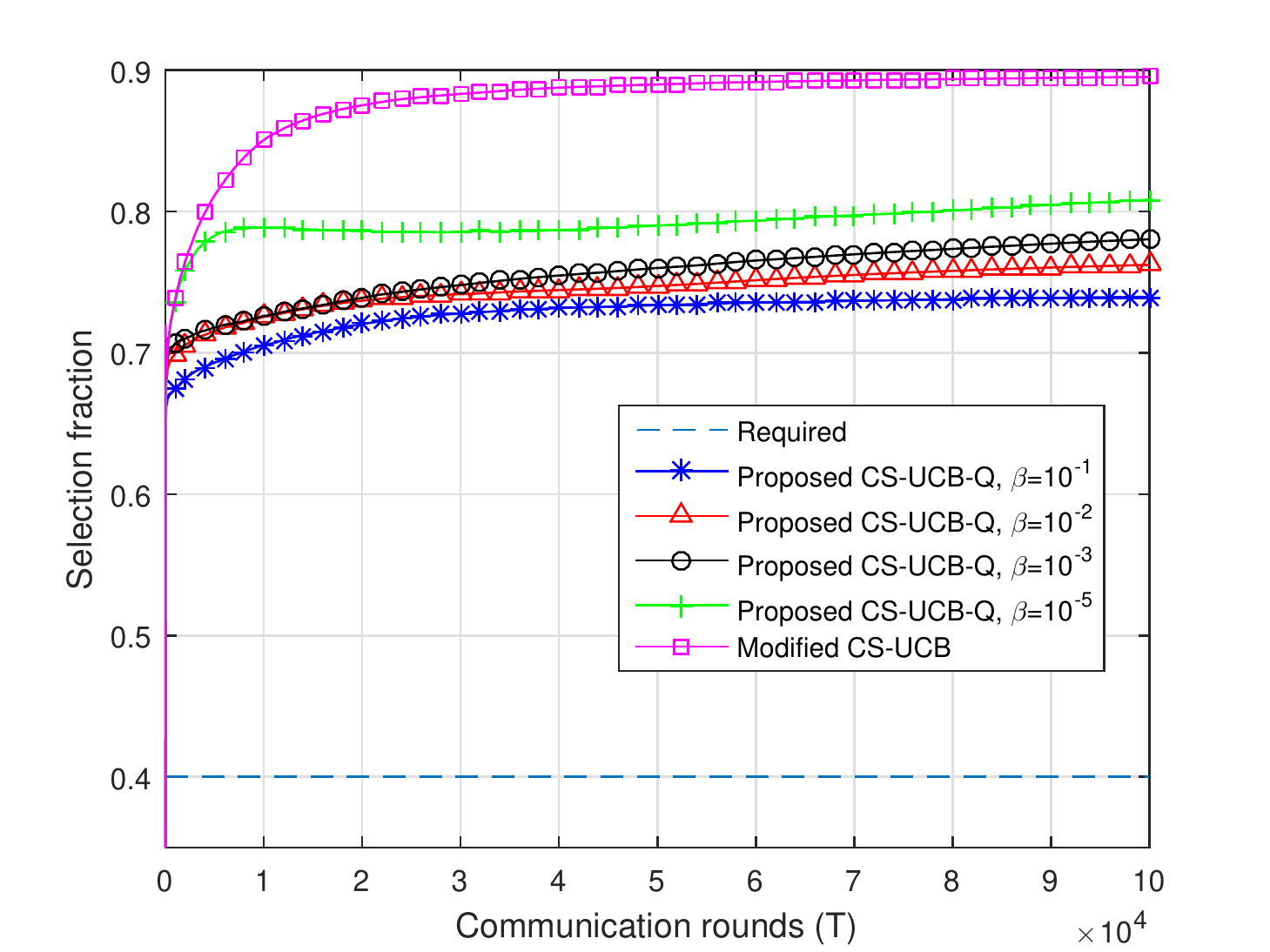}}
  \caption{Selection fractions of different clients versus the number of communication rounds with \{$K=3$, $N=2$\}.}
  \label{arms}
\end{figure}

Before showing the performance of the proposed CS-UCB-Q algorithm,  a baseline algorithm based on the proposed CS-UCB algorithm is introduced for comparison, which takes the availability of clients into account but does not consider the fairness constraint. Thus, we refer to such a baseline algorithm as modified CS-UCB algorithm.
The cumulative  and average performance gains are given as $\Theta=\tau_{\max}\Sigma^{\pi_2}$ and $\theta=\Theta/T$, respectively, where $\pi_2$ can be the proposed CS-UCB-Q algorithm or the  modified CS-UCB algorithm.
We consider a system with $K=3$ and $N=2$, where the fairness constraint vector is set as $\qc=(c_1,c_2,c_3)=(0.6,0.5,0.4)$ and the sample vector is $\mathbf{s}=(s_1,s_2,s_3)=(1500,1000,500)$. Besides, the availability of each client $k$ is  a binary random variable  with  mean $\alpha_k$ which is i.i.d. over time. For simplicity, we set $\alpha_k=0.9,\forall k\in\calK$.

In Fig. \ref{comparison_round_ns}, different $\beta$ values are considered and these $\beta$ values make the proposed CS-UCB-Q algorithm meet the fairness constraint. Different from the proposed CS-UCB-Q algorithm, the modified CS-UCB algorithm is unaware of the fairness constraint and thus can achieve the smallest performance gap, which is even negative, i.e., it achieves better performance than the optimal solution obtained by solving the linear program \eqref{linear problem}.  Besides, we also find that a smaller $\beta$ value in the proposed CS-UCB-Q algorithm results in a smaller performance gap since the importance of the fairness constraint is reduced and more priority is given to the reward of each client. Such an observation at first glance seems to suggest that the proposed CS-UCB-Q algorithm with a smaller $\beta$ value is more attractive as it achieves a better performance while satisfying the fairness constraint. However, the speed of convergence to a point that meets the fairness constraint is not reflected in Fig. \ref{comparison_round_ns}, which is also an important concern in practice.

Fig. \ref{arms} presents the selection fractions of different clients with respect to the number of  communication rounds. We find that the curve with a smaller $\beta$ value has a slow  speed of convergence to the point satisfying the fairness constraint. For example,  the convergence rate of the proposed CS-UCB-Q algorithm with $\beta=10^{-5}$ is the slowest and the fairness constraint of client 1 is violated in the whole learning process. Thus, there is a tradeoff between the performance and the convergence rate. Besides, we can infer from Fig. \ref{arms} that the reward of client 1 is the lowest and that of client 3 is the highest. In other words, the time consumed by the client 1 to finish its local training and transmission is the longest and that of client 3 is the shortest. This is because, without the limitation of fairness, the modified CS-UCB algorithm tends to choose the fast clients to participate in the FL process.

\begin{figure}
      \begin{minipage}[t]{0.47\textwidth}
        \centering
        \includegraphics[width=1\textwidth]{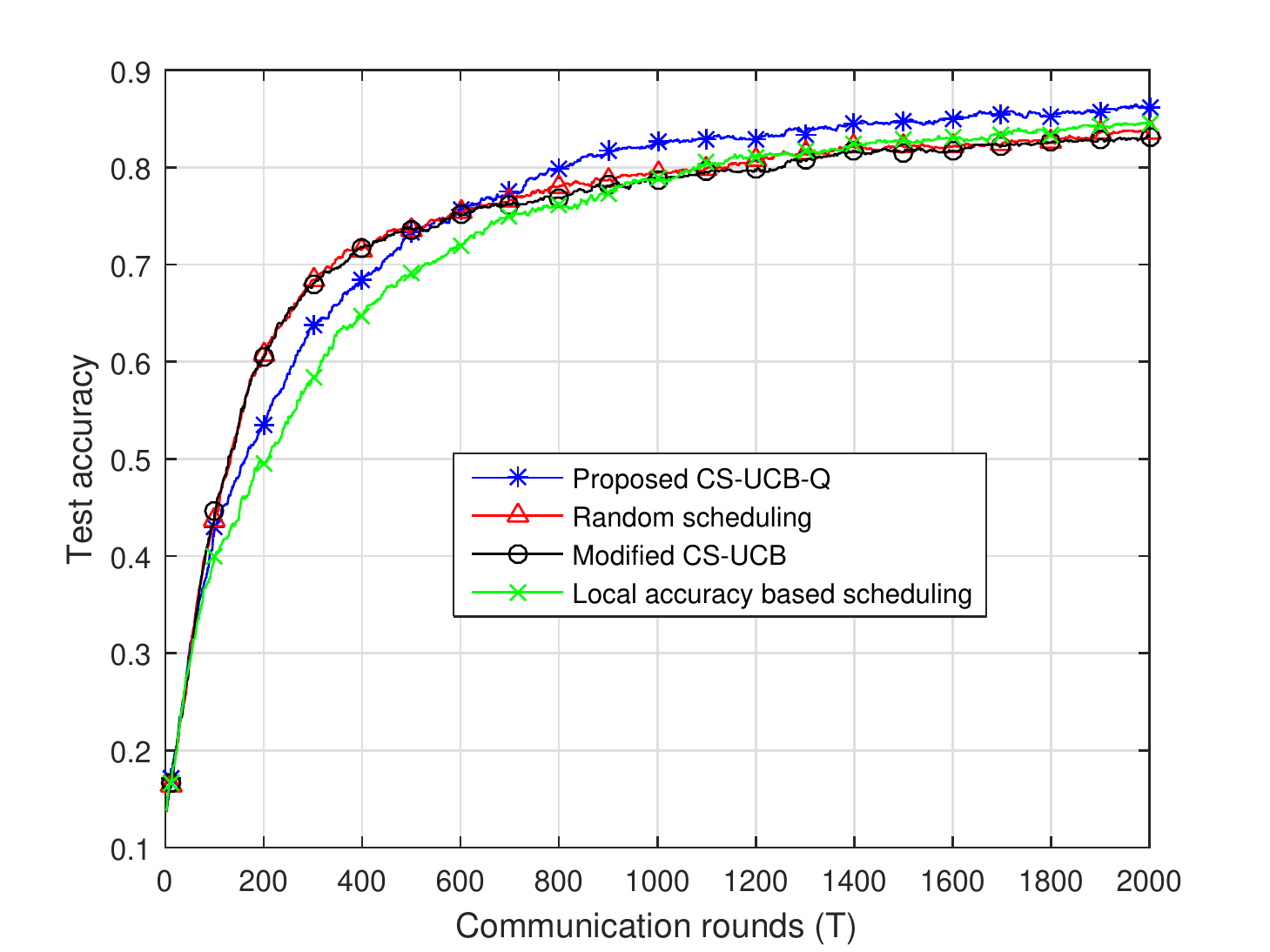}
        \caption{Test accuracy versus the number of communication rounds of the proposed CS-UCB-Q algorithm and baselines with \{$K=10,N=4$\}.}
        \label{test_acc_nonideal}
      \end{minipage}%
      \hfill
      \begin{minipage}[t]{0.47\textwidth}
        \centering
        \includegraphics[width=1\textwidth]{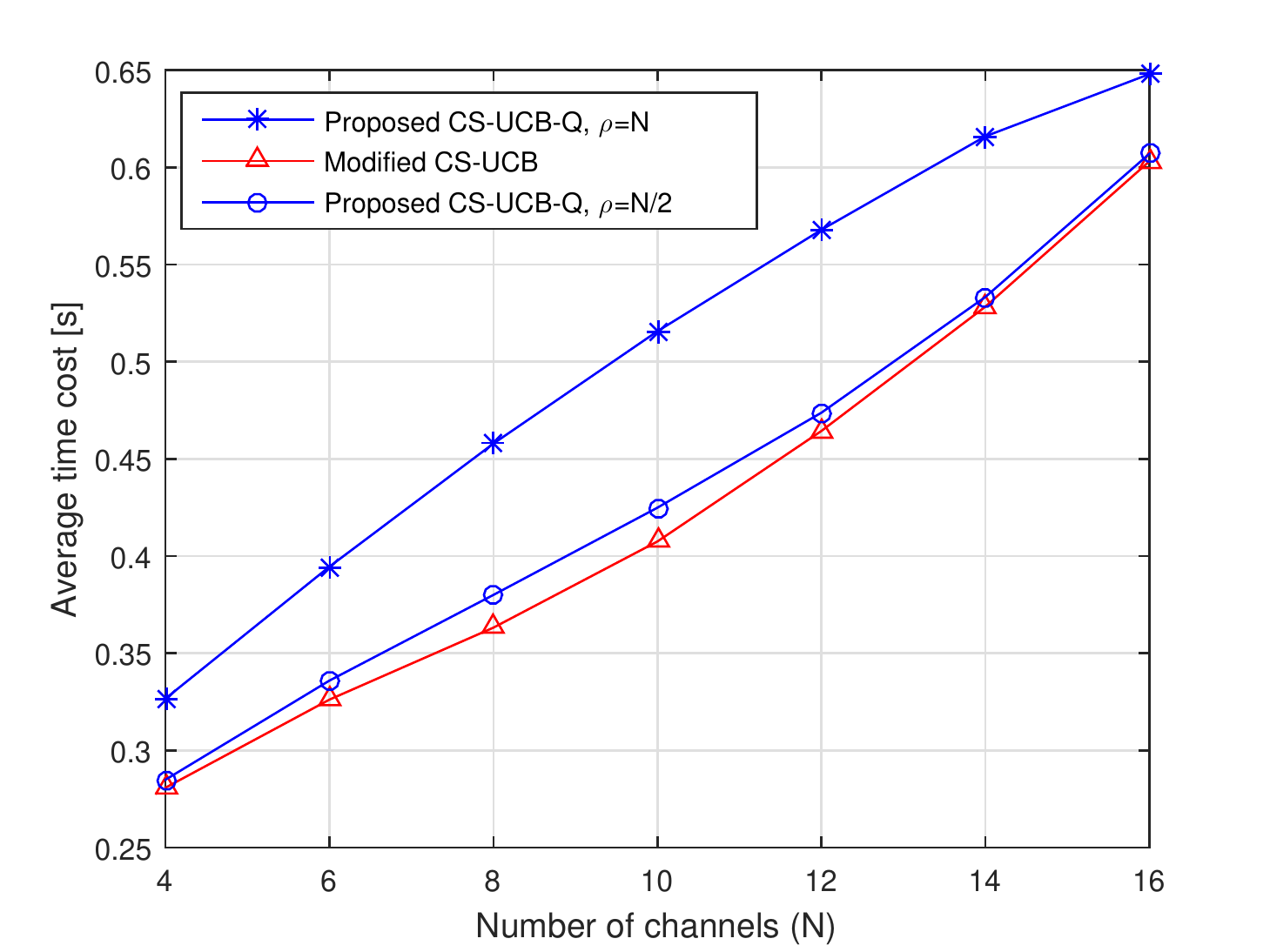}
        \caption{Performance comparison of different algorithms with respect to the number of channels with \{$T=2\times 10^4$, $K=20$\}.}
        \label{cum_gap_n_ns}
      \end{minipage}
    \end{figure}

To simulate a heterogeneous setting in the non-ideal scenario, we distribute the training samples among $K=10$ clients such that two clients have  samples of 5 digits and the other clients have  samples of only 1 digit.  We  assume that the fairness factor is proportional to the size of local dataset, i.e., $c_k=\frac{N}{2}\frac{s_k}{\sum_{k\in\calK}s_k}$ with $N=4$.  The test accuracy with respect to the number of communication rounds of the proposed CS-UCB-Q algorithm, as well as the baseline algorithms, is presented in Fig. \ref{test_acc_nonideal}.  It is shown that the achieved test accuracy of the proposed CS-UCB-Q algorithm is better than that of the three baseline algorithms. This is because the proposed CS-UCB-Q algorithm  makes the clients with the important datasets participate in more rounds.  According to the results of Figs. \ref{comparison_round_ns}(a) and \ref{test_acc_nonideal},  we find that the proposed CS-UCB-Q algorithm can reduce the training latency without degrading the learning performance.  In other words, given the number of communication rounds required to achieve a certain level of test accuracy, the proposed CS-UCB-Q algorithm can finish the FL training process with less wall-clock time.

To show the performance with respect to the number of available channels $N$, we assume that the fairness constraint is proportional to the size of the local dataset of each client, i.e., $c_k=\rho\frac{s_k}{\sum_{k\in\calK}s_k}$, where $\rho$ is a scaling factor. As shown in Fig. \ref{cum_gap_n_ns}, the larger $\rho$ value leads to a larger delay. This is because  the larger $\rho$ value suggests that more communication rounds are forced to meet the requirement for the fairness constraint and thus the slow clients  can participate in more communication rounds,   leading to more time consumption.

\vspace{-0.3cm}
\section{Conclusions}\label{section conclusions}
Aiming to minimize the time consumption of FL training, this work considered the CS problem in both the ideal scenario and non-ideal scenarios. For the ideal scenario, we proposed the CS-UCB algorithm and also derived an upper bound of its performance regret. The upper bound suggests that the performance regret of the proposed CS-UCB algorithm  grows in a logarithmic way over communication rounds. However, the local datasets of clients are non-i.i.d. and unbalanced   and  the availability of clients is dynamic in the non-ideal scenario. Thus, we introduced the fairness constraint to ensure each client could participate in a certain proportion of the communication rounds during  the training process. We further proposed the  CS-UCB-Q algorithm based on UCB policy  and virtual queue technique and provided an upper bound which shows that the performance regret of the proposed CS-UCB-Q algorithm has a sub-linear growth over communication rounds when $\beta\leq\frac{1}{\sqrt{T}}$ and $T$ is large enough. Simulation results validate the efficiency of the proposed CS-UCB and CS-UCB-Q algorithms and also reveal a tradeoff between the performance regret and the speed of convergence to a point where the fairness constraint is  satisfied.

\appendices
\vspace{-0.3cm}
\section{Proof of Theorem \ref{upper bound theorem 1}}\label{proof of upper bound 1}
Based on \eqref{regret 1}, we have the following inequality
\begin{equation}\label{eq xl}
\begin{split}
\Sigma^{\pi_1}
&=T\min_{k\in\calS^{\ast}}\mu_k-\sum_{t=1}^{T}\Ex\big[\min_{k\in\calS(t)}\mu_k\big]\leq T\sum_{k\in\calS^{\ast}}\mu_k-\sum_{t=1}^{T}\Ex\big[\sum_{k\in\calS(t)}\mu_k\big]\\
&=\sum_{\calS:r(\calS)\leq r(\calS^{\ast})}\Delta_{\calS}\Ex[v_{\calS}(T)]\leq  \Delta_{\max}\sum_{\calS:r(\calS)\leq r(\calS^{\ast})}\Ex[v_{\calS}(T)],
\end{split}
\end{equation}
where $v_{\calS}(T)$ is the number of times the super arm $\calS$ is chosen in the first $T$ rounds. Then, a counter vector $\qn\in\mathbb{R}^{K\times1}$ is  introduced. In each round after the initialization period, one of the two following cases must occur: 1) the optimal super arm is selected; 2) a non-optimal super arm is selected. $\qn$ is not updated in the first case.  In the second case where a non-optimal super arm $\calS(t)$ is selected, the counter $n_k$ with $k=\argmin_{k^{\prime}}z_{k^{\prime}}$ is increased by 1 if $k$ is unique. However, if more than one arms satisfy  $k=\argmin_{k^{\prime}}z_{k^{\prime}}$,  of such arms a random one is chosen and its counter is increased by 1. Thus, the total number of the non-optimal actions are  played has the following constraint:
\begin{equation}\label{eq x2}
  \sum_{\calS:r(\calS)\leq r(\calS^{\ast})}v_{\calS}(t)\leq \kappa+\sum_{k\in\calK}n_k(t), t\geq \kappa+1.
\end{equation}
Based on \eqref{eq xl} and \eqref{eq x2}, we have
\begin{equation}\label{eq x3}
\Sigma^{\pi_1}\leq  \Delta_{\max}\{\kappa+\sum_{k\in\calK}\Ex[n_k(T)]\}.
\end{equation}

Now, we focus on the expectation of $n_k(T), \forall k\in\calK$. We define $l_k$ as the indicator function which is equal to 1 if $n_k(t)$ is added by 1 in round $t$ and also define $\mathbb{I}(x)$ as the indicator function which is equal to 1 if $x$ is true; $\mathbb{I}(x)=0$ otherwise. Let $f_{t,z_k}=\sqrt{(N+1)\ln t/ z_k}$ and  $\bar{y}_{k,z_k}$ denote the sample mean of all the observed values of $r_k$ when it is observed $z_k$ times. Then,
\begin{equation}
\begin{split}
  n_k(T)
  &\leq l+\sum_{t=\kappa}^T\mathbb{I}\Big\{\sum_{k\in\calS^{\ast}}\bar{y}_{k,z_k(t)}+f_{t},z_k(t) \leq\sum_{k\in\calS(t+1)}\bar{y}_{k,z_k(t)}+f_{t,z_k(t)},n_k(t)\geq l \Big\}.
\end{split}
\end{equation}
Since $l\leq n_k(t)<z_k(t)$, we further have
\begin{equation}
\begin{split}
  n_k(T) 
  &\leq l+\sum^{\infty}_{t=1}\sum^{t}_{\tilde{k}_1=1}\cdots\sum^{t}_{\tilde{k}_N=1}\sum^{t}_{\hat{k}_1=l}\cdots\sum^{t}_{\hat{k}_N=l}\mathbb{I}\Big\{\sum_{j=1}^N\bar{y}_{\tilde{k}_j,z_{\tilde{k}_j}}+f_{t,z_{\tilde{k}_j}} \leq\sum^N_{j=1}\bar{y}_{\hat{k}_j,z_{\hat{k}_j}}+f_{t,z_{\hat{k}_j}}\Big\},
\end{split}
\end{equation}
where $\tilde{k}_j$ is the $j$-th element of $\calS^{\ast}$ and  $\hat{k}_j$ is the $j$-th element of $\calS(t+1)$. Note that $\sum_{j=1}^N\bar{y}_{\tilde{k}_j,z_{\tilde{k}_j}}+f_{t,z_{\tilde{k}_j}} \leq\sum^N_{j=1}\bar{y}_{\hat{k}_j,z_{\hat{k}_j}}+f_{t,z_{\hat{k}_j}}$ suggests that at least one of the following inequalities must hold:
\begin{subequations}
\begin{align}
\sum\nolimits_{j=1}^N\bar{y}_{\tilde{k}_j,z_{\tilde{k}_j}} &\leq     \sum\nolimits^N_{j=1}\mu_{\tilde{k}_j}- \sum\nolimits_{j=1}^Nf_{t,z_{\tilde{k}_j}},\label{event1} \\
  \sum\nolimits^N_{j=1}\bar{y}_{\hat{k}_j,z_{\hat{k}_j}} & \geq   \sum\nolimits^N_{j=1}\mu_{\hat{k}_j}+ \sum\nolimits^N_{j=1}f_{t,z_{\hat{k}_j}},\label{event2} \\
 \sum\nolimits^N_{j=1}\mu_{\tilde{k}_j}&<  \sum\nolimits^N_{j=1}\mu_{\hat{k}_j}+ 2\sum\nolimits^N_{j=1}f_{t,z_{\hat{k}_j}}.\label{event3}
\end{align}
\end{subequations}
We bound the probability of inequality  \eqref{event1} as
\begin{equation}
\begin{split}
\Pr\{\eqref{event1} \ \text{holds}\}
  &\overset{(i)}{\leq}\sum_{j=1}^N e^{-2(N+1)\ln t}= N t^{-2(N+1)},
 \end{split}
\end{equation}
where (i) holds because of the Chernoff-Hoeffding bound \cite{pollard2012convergence}.   Similarly, the probability of inequality \eqref{event2} is given as $\Pr\{\eqref{event2} \ \text{holds}\}\leq  N t^{-2(N+1)}$.
Note that for $l\geq\lceil \frac{4(N+1)N^2\ln T }{(\Delta_{\calS(t+1)})^2}\rceil$, we have $\sum^N_{j=1}\mu_{\tilde{k}_j}-  \sum^N_{j=1}\mu_{\hat{k}}- 2\sum^N_{j=1}f_{t,z_{\hat{k}_j}}\geq0$,
which suggests that \eqref{event3} does not hold if $l\geq \lceil \frac{4(N+1)N^2\ln T }{(\Delta_{\calS(t+1)})^2}\rceil$. Based on this fact, we let $l= \lceil \frac{4(N+1)N^2\ln T }{(\Delta_{\min})^2}\rceil$  with $\Delta_{\min}=\min_{r(\calS)\leq r(\calS^{\ast})} \Delta_{\calS}$ and then have
\begin{equation}\label{eq x4}
 \begin{split}
\Ex[n_k(T)]&\leq\lceil \frac{4(N+1)N^2\ln T }{(\Delta_{\min})^2}\rceil+\sum^{\infty}_{t=1}\sum^{t}_{\tilde{k}_1=1}\cdots\sum^{t}_{\tilde{k}_N=1}\sum^{t}_{\hat{k}_1=l}\cdots\sum^{t}_{\hat{k}_1=l}2N t^{-2(N+1)}\\
&\leq \frac{4(N+1)N^2\ln T }{(\Delta_{\min})^2}+1 +\frac{\pi^2}{3}N.
\end{split}
\end{equation}
Finally, the proof is finished by substituting \eqref{eq x4} into \eqref{eq x3} with the fact $\kappa< K/N+1$.

\vspace{-0.5cm}
\section{Proof of Theorem \ref{convergence theorem 1}}\label{proof of convergence theorem 1}
According to \textbf{Assumption \ref{L-smoothness}} and taking $\gamma\leq\frac{1}{L}$, we have
\begin{align}
  G(\qx(t+1))-  G(\qx(t))&\leq -\gamma  \nabla G^{\dag}(\qx(t))\qv(t)+\frac{\gamma}{2}||\qv(t)||_2^2\nonumber \\
  &= -\frac{\gamma}{2}||\nabla G(\qx(t))||_2^2+\frac{\gamma}{2}||\nabla G(\qx(t))-\qv(t)||_2^2.
\end{align}
We further take expectation over $\qv(t)$ and then obtain
\begin{equation}\label{qq1}
\begin{split}
  &\Ex\big[G(\qx(t+1))\big]-  G(\qx(t))\\
  \leq& -\frac{\gamma}{2}||\nabla G(\qx(t))||_2^2+\frac{\gamma}{2}\Ex\bigg\{\Big|\Big|\frac{1}{N}\sum_{k\in\calS(t)}\Big[\nabla G(\qx(t))-\nabla G_k(\qx(t))\Big]\Big|\Big|_2^2\bigg\} \\
  \leq &-\frac{\gamma}{2}||\nabla G[\qx(t)]||_2^2+\frac{\gamma \delta_0}{2N},
\end{split}
\end{equation}
where the last inequality holds because of \textbf{Assumption \ref{Bounded variance}} and $\Ex[\nabla G_k(\qx(t))]=G(\qx(t))$. Now, we define a new function  $\mathcal{G}(\hat{\qx})=G(\qx(t))+\nabla G^{\dag}(\qx(t))(\hat{\qx}-\qx(t))+\frac{\Phi}{2}||\hat{\qx}-\qx(t)||^2_2$, whose minimal value is achieved when all the partial  derivatives are 0's, i.e., $\frac{\partial\mathcal{G}(\hat{\qx})}{\partial \hat{\qx}}=\nabla G(\qx(t))+\Phi[\hat{\qx}(t)-\qx(t)]=\bm{0}$. The optimal point is  $\hat{\qx}^{\ast}=\qx(t)-\frac{\nabla G(\qx(t))}{\Phi}$ and the minimal value is $\mathcal{G}_{\min}=G(\qx(t))-\frac{||\nabla G(\qx(t))||^2}{2\Phi}$. According to \textbf{Assumption \ref{Strong convexity}}, we obtain
\begin{equation}\label{qq0}
  G(\qx^{\ast})\geq \mathcal{G}(\qx^{\ast})\geq  \mathcal{G}_{\min},
\end{equation}
which suggests that
\begin{equation}\label{qq2}
 2\Phi[G(\qx(t))-G(\qx^{\ast})]\leq ||\nabla G(\qx(t))||_2^2.
\end{equation}
Combining \eqref{qq1} and \eqref{qq2} leads to
\begin{equation}\label{qq3}
\Ex[G(\qx(t+1))]-G(\qx(t))\leq  -\gamma\Phi[G(\qx(t))-G(\qx^{\ast})]+\frac{\gamma\delta_0}{2N}.
\end{equation}
We rearrange \eqref{qq3} as
\begin{equation}\label{qq4}
\Ex[G(\qx(t+1))]-G(\qx^{\ast})\leq  (1-\gamma\Phi)[G(\qx(t))-G(\qx^{\ast})]+\frac{\gamma\delta_0}{2N}.
\end{equation}
Then, taking total expectations over $\{\qv(t)\}^\dag_{t=1}$ and subtracting a constant $\frac{\delta_0}{2N\Phi}$ from both sides of \eqref{qq4}, we have
\begin{equation}\label{qq5}
 \Ex\bigg[G(\qx(t+1))]-G(\qx^{\ast})-\frac{\delta_0}{2N\Phi}\bigg]\leq  (1-\gamma\Phi)\Ex\bigg[G(\qx(t))]-G(\qx^{\ast})-\frac{\delta_0}{2N\Phi}\bigg],
\end{equation}
which is a contraction inequality because of $0<\gamma\Phi\leq 1$. Finally, the result in \textbf{Theorem \ref{convergence theorem 1}} can be obtained by applying \eqref{qq5} repeatedly.
\vspace{-0.5cm}
\section{Proof of Theorem \ref{feasibility optimality theorem}}\label{proof of feasibility optimality theorem}
To validate \textbf{Theorem \ref{feasibility optimality theorem}},  we need to  prove that all the virtual queues defined in  \eqref{queue} is mean rate stable \cite[Definition 2.2] {neely2010stochastic}, i.e.,
\begin{equation}
  \lim_{T\rightarrow\infty}\frac{\Ex[\sum^K_{k=1}D_k(T)]}{T}=0.
\end{equation}
We first introduce the  Lyapunov function $L(\qD(t))=\frac{1}{2}\sum_{k\in\calK}D^2_k(t)$,
of which the drift is given as
\begin{equation}\label{inequ1}
\setlength{\abovecaptionskip}{0.cm}
\setlength{\belowcaptionskip}{-0.cm}
\begin{split}
L(\qD(t+1))-L(\qD(t))
&\leq \frac{1}{2}\sum_{k\in\calK}\max\{c_k^2,(1-c_k)^2\}+\sum_{k\in\calK}(c_k-b_k(t))D_k(t)\\
&= \frac{\Omega}{2}+\sum_{k\in\calK}(c_k-b_k(t))D_k(t).
\end{split}
\end{equation}
Then, we take conditional expectation of both sides of the inequality \eqref{inequ1} and the resulting conditional  Lyapunov function is  given as
\begin{equation}\label{eq drift}
\begin{split}
\Ex[L(\qD(t+1))-L(\qD(t))|\qD(t)]&\leq \frac{\Omega}{2}+\sum_{k\in\calK}c_kD_k(t)-\Ex[\sum_{k\in\calK}b_k(t)D_k(t)|\qD(t)]\\
&\leq\Xi+\sum_{k\in\calK}c_kD_k(t)-\frac{1}{\beta}\Ex[\sum_{k\in\calS(t)}\upsilon_k(t)|\qD(t)],
\end{split}
\end{equation}
where  $\Xi=\frac{\Omega}{2}+\frac{(1-\beta)N}{\beta}$ and $\upsilon_k(t)=(1-\beta)\hat{y}_k(t)+\beta D_k(t)$. The last inequality holds because $\hat{y}_k(t)\leq1,\forall k\in\calK$. Since $\mathbf{c}$ is assumed to be strictly in $\calC$, there must be $\epsilon>0$ such that  $\mathbf{c}+\epsilon\mathbf{1}$ is also strictly in $\calC$, where $\mathbf{1}=[1,1,\ldots,1]^T\in\mathbb{R}^{K\times 1}$. According to \textbf{Lemma \ref{lemma 1}}, one can always find an $\calA$-only policy $\pi_2^{\prime}$ which can support the minimum selection fraction vector  $\mathbf{c}+\epsilon\mathbf{1}$, i.e.,
\begin{equation}\label{eq fraction}
\Ex[b^{\prime}_k(t)]=\sum_{e\in\calP(\calK)}\hat{P}_{\calA}(e)\sum_{\calS\in\calQ(e):k\in \calS}q^{\prime}_{\calS}(e)\geq c_k+\epsilon, \forall k\in\calK.
\end{equation}
where $b^{\prime}_k(t)$ and $q^{\prime}_{\calS}(e)$ are the corresponding variables under policy $\pi_2^{\prime}$.
Define $\calS^{\prime}(t)$ as the super arm chosen by policy $\pi_2^{\prime}$ at round $t$.  Then, the lower bound of $\Ex[\sum_{k\in\calS(t)}\upsilon_k(t)|\qD(t)]$ is given by
\begin{equation}\label{eq 11}
\begin{split}
&\Ex\Big[\sum_{k\in\calS(t)}\upsilon_k(t)|\qD(t)\Big]=\Ex\Big[\Ex\Big[\sum_{k\in\calS(t)}\upsilon_k(t)|\qD(t),\calA(t)\Big]\Big]\\
\overset{(i)}{\geq}&\Ex\Big[\Ex\Big[\sum_{k\in\calS^{\prime}(t)}\upsilon_k(t)|\qD(t),\calA(t)\Big]\Big]\geq \beta \Ex\Big[\Ex\Big[\sum_{k\in\calS^{\prime}(t)}D_k(t)|\qD(t),\calA(t)\Big]\Big]\\
\overset{(ii)}{=} &\beta \Ex\Big[\Ex\Big[\sum_{k\in\calS^{\prime}(t)}D_k(t)|\calA(t)\Big]\Big]
=\beta \sum_{k\in\calK}D_k(t)\sum_{e\in\calP(\calK)}\hat{P}_{\calA}(e)\sum_{\calS\in\calQ(e):k\in \calS}q^{\prime}_{\calS}(e),
\end{split}
\end{equation}
in which (i) holds because of the expression \eqref{arm selection} and (ii) holds due to the independence of policy $\pi_2^{\prime}$ on $\qD(t)$.
Applying \eqref{eq fraction} to \eqref{eq 11}, we can obtain
\begin{equation}\label{eq 22}
 \Ex\Big[\sum_{k\in\calS(t)}\upsilon_k(t)|\qD(t)\Big]\geq \beta \sum_{k\in\calK}D_k(t)( c_k+\epsilon), k\in\calK.
\end{equation}
Next, we substitute \eqref{eq 22} into \eqref{eq drift} and have
\begin{equation}\label{eq 33}
\begin{split}
\Ex[L(\qD(t+1))-L(\qD(t))|\qD(t)]
&\leq \Xi+\sum_{k\in\calK}c_kD_k(t)- \sum_{k\in\calK}D_k(t)(c_k+\epsilon)\\
&= \Xi-\epsilon\sum_{k\in\calK}D_k(t).
\end{split}
\end{equation}
According to the Lyapunov drift theorem \cite[Theorem 4.1]{neely2010stochastic} and given that  $\epsilon>0$,  the result in \eqref{eq 33} suggests that all the virtual queues defined in  \eqref{queue} is not only mean rate stable, but also strongly stable \cite[Definition 2.7]{neely2010stochastic} , i.e.,
\begin{equation}
  \lim_{T\rightarrow\infty}\sup\frac{1}{T}\sum_{t=1}^T\Ex\left[D_k(t)\right]\leq\frac{\Xi}{\epsilon}<\infty.
\end{equation}
Thus, the proof is completed.
\vspace{-0.5cm}
\section{Proof of Theorem \ref{upper bound theorem 2}}\label{proof of upper bound theorem 2}
Given the optimal $\calA$-policy $\pi_2^\ast$, we denote by $\calS^{\ast}(t)$ the  super arm chosen in round $t$ by policy $\pi_2^\ast$ and $b_k^{\ast}(t)$'s are the indicator variables that correspond to $\calS^{\ast}(t)$.  Furthermore, we define $\Upsilon_1(t)=\sum_{k\in\calK}\big(b_k^{\ast}(t)-b_k(t)\big)\mu_k$
and $\Upsilon_2(t)=\sum_{k\in\calK}[\beta D_k(t)+(1-\beta)\mu_k]\big(b^{\ast}_k(t)-b_k(t)\big)$. The following lemma gives a related inequality between  $\Upsilon_1(t)$ and  $\Upsilon_2(t)$.
\begin{lemma}\label{lemma 2}
The sum of the time sequence $\Ex[\Upsilon_1(t)], t=1,2,\ldots,T$, is bounded by
\begin{equation}\label{lemma2_result}
\begin{split}
\sum^T_{t=1}\Ex[\Upsilon_1(t)]&\leq\frac{\beta\Omega T}{2}+ \sum^T_{t=1}\Ex\left[\Upsilon_2(t)\right].
\end{split}
\end{equation}
\end{lemma}
\begin{IEEEproof}
To validate \textbf{Lemma \ref{lemma 2}}, we first introduce a variable $\xi(t)=\beta[L(\qD(t+1))-L(\qD(t)]+(1-\beta)\Upsilon_1(t)$ which is referred to as the drift-plus-regret. The upper bound of $\Ex[\xi(t)]$ is given as
\begin{equation}
\begin{split}
\Ex[\xi(t)]&\leq\beta\Ex\Big[\frac{\Omega}{2}+\sum_{k\in\calK}(c_k-b_k(t))D_k(t)\Big]+(1-\beta)\Ex \Big[\sum_{k\in\calK} b_k^{\ast}(t)\mu_k-\sum_{k\in\calK}b_k(t)\mu_k\Big]\\
&=\frac{\beta\Omega}{2}+\Ex[\Upsilon_2(t)]+\beta\sum_{k\in\calK} \Ex [ D_k(t)(c_k-b^{\ast}_k(t))]\leq\frac{\beta\Omega}{2}+\Ex[\Upsilon_2(t)],
\end{split}
\end{equation}
where the last inequality holds because of \eqref{fairness constraint}. By summing  $\Ex[\xi(t)]$ over $t=1,2\ldots, T$, we obtain
\begin{equation}\
\begin{split}
  \sum^T_{t=1}\Ex[\xi(t)] &=\beta\Ex[L(\qD(T+1))-L(\qD(1)]+\sum^T_{t=1}\Ex[\Upsilon_1(t)]\leq\frac{\beta\Omega T}{2}+ \sum^T_{t=1}\Ex\left[\Upsilon_2(t)\right].
\end{split}
\end{equation}
Since $L(\qD(1))=0$ and $L(\qD(T+1))\geq 0$, we obtain the inequality \eqref{lemma2_result} directly. The proof of  \textbf{Lemma \ref{lemma 2}} is completed.
\end{IEEEproof}

With the result in \textbf{Lemma \ref{lemma 2}} and the following inequality
\begin{equation}
\begin{split}
\Sigma^{\pi_2}
&=\sum_{t=1}^{T}\Ex\left[\mu(\calS^{\ast}(t))-\mu(\calS(t))\right]=\sum_{t=1}^{T}\Ex\Big[\min_{k\in\calS^{\ast}(t)}\mu_k-\min_{k\in\calS(t)}\mu_k\Big]\\
&\leq\sum_{t=1}^{T}\Ex\Big[\sum_{k\in\calS^{\ast}(t)}\mu_k-\sum_{k\in\calS(t)}\mu_k\Big]=\sum_{t=1}^{T}\Ex[\Upsilon_1(t)],
\end{split}
\end{equation}
the expected regret is bounded by
\begin{equation}\label{sub_result1}
\Sigma^{\pi_2}\leq\frac{\beta\Omega T}{2}+\sum^T_{t=1}\Ex\left[\Upsilon_2(t)\right].
\end{equation}
To give the upper bound of $\sum^T_{t=1}\Ex\left[\Upsilon_2(t)\right]$, we define a super arm $\calS^{\ddag}(t)$ chosen by another policy $\pi_2^{\ddag}$ in each round $t$ with the following rule:
\begin{equation}
  \calS^{\ddag}(t)\in\mathop{\argmax}_{\calS\in\calQ(\calA(t))}\sum_{i\in\calS}\beta D_k(t)+(1-\beta)\mu_k.
\end{equation}
Given $\calS(t)$ is chosen according to \eqref{arm selection},  then we directly have the following inequality
\begin{equation}\label{ttep}
 \sum_{i\in\calS^{\ddagger}(t)}(1-\beta)\hat{y}_k(t)+\beta D_k(t) \leq \sum_{k\in\calS(t)} (1-\beta)\hat{y}_k(t)+\beta D_k(t).
\end{equation}
Based on \eqref{ttep}, we bound $\Upsilon_2(t)$ by
\begin{equation}\label{sub_result2}
  \begin{split}
  \Upsilon_2(t)
  &\leq \sum_{k\in\calS^{\ddag}(t)}[\beta D_k(t)+(1-\beta)\mu_k]-\sum_{k\in\calS(t)}[\beta D_k(t)+(1-\beta)\mu_k]\\
  &+\sum_{k\in\calS(t)} [\beta D_k(t)+(1-\beta)\hat{y}_k(t)]-\sum_{i\in\calS^{\ddag}(t)}[\beta D_k(t)+(1-\beta)\hat{y}_k(t)]\\
  &=(1-\beta)[\Upsilon_3(t)+\Upsilon_4(t)],
  \end{split}
\end{equation}
where $\Upsilon_3(t)=\sum_{k\in\calS(t)} \big(\hat{y}_k(t)-\mu_k\big)$ and $\Upsilon_4(t)=\sum_{i\in\calS^{\ddag}(t)}\big(\mu_k-\hat{y}_k(t)\big)$.
Here, we directly give the upper bounds of $\Lambda_1=\sum_{t=1}^T\Ex [\Upsilon_3(t)]$ and $\Lambda_2=\sum_{t=1}^T\Ex [\Upsilon_4(t)]$ as follows:
\begin{equation}\label{sub_result3}
\Lambda_1\leq  (\frac{\pi^2}{6}+1)K+4\sqrt{2KNT\ln T} \ \text{and} \ \Lambda_2\leq \frac{\pi^2}{6}K.
\end{equation}
The related  analysis follows a similar line of that in \cite{li2019combinatorial}.

Finally, by plugging  \eqref{sub_result3} into \eqref{sub_result2} and further  into \eqref{sub_result1}, the proof is completed.

\vspace{-0.5cm}
\section{Proof of Theorem \ref{convergence theorem 2}}\label{proof of convergence theorem 2}
According to \textbf{Assumption \ref{L-smoothness}} and taking expectation on $\qv(t)$, we have
\begin{equation}
\Ex[G(\qx(t+1))]-G(\qx(t))\leq -\gamma \nabla G^{\dag}(\qx(t))\Ex[\qv(t)]+\frac{L\gamma^2}{2}\Ex||\qv(t)||_2^2,
\end{equation}
where $\Ex||\qv(t)||_2^2$ is bounded by, based on \textbf{Assumptions \ref{The first and second moment conditions}(b)} and \textbf{5(c)},
\begin{align}
\Ex||\qv(t)||_2^2 &\leq ||\Ex[\qv(t)]||_2^2+\delta_3+||\nabla G(\qx(t))||_2^2 \leq \delta_3+(1+\delta^2_1)||\nabla G(\qx(t))||_2^2.
\end{align}
Thus,
\begin{align}
 \Ex[G(\qx(t+1))]-G(\qx(t))&\leq -\gamma[\delta_2-\frac{L\gamma}{2}(1+\delta^2_1)]||\nabla G(\qx(t))||_2^2+\frac{L\gamma^2\delta_3}{2}\nonumber \\
 &\leq -\frac{\gamma\delta_2}{2}||\nabla G(\qx(t))||_2^2+\frac{L\gamma^2\delta_3}{2},
\end{align}
where the second inequality holds because of $\gamma\leq\frac{\delta_2}{L(1+\delta_1^2)}$. Finally, we take similar actions in \eqref{qq0}-\eqref{qq4} and the proof is completed.

\vspace{-0.7cm}

\begin{thebibliography}{10}
\providecommand{\url}[1]{#1}
\csname url@samestyle\endcsname
\providecommand{\newblock}{\relax}
\providecommand{\bibinfo}[2]{#2}
\providecommand{\BIBentrySTDinterwordspacing}{\spaceskip=0pt\relax}
\providecommand{\BIBentryALTinterwordstretchfactor}{4}
\providecommand{\BIBentryALTinterwordspacing}{\spaceskip=\fontdimen2\font plus
\BIBentryALTinterwordstretchfactor\fontdimen3\font minus
  \fontdimen4\font\relax}
\providecommand{\BIBforeignlanguage}[2]{{%
\expandafter\ifx\csname l@#1\endcsname\relax
\typeout{** WARNING: IEEEtran.bst: No hyphenation pattern has been}%
\typeout{** loaded for the language `#1'. Using the pattern for}%
\typeout{** the default language instead.}%
\else
\language=\csname l@#1\endcsname
\fi
#2}}
\providecommand{\BIBdecl}{\relax}
\BIBdecl

\bibitem{mcmahan2016communication}
H.~B. McMahan, E.~Moore, D.~Ramage, S.~Hampson \emph{et~al.},
  ``Communication-efficient learning of deep networks from decentralized
  data,'' \emph{arXiv preprint arXiv:1602.05629}, 2016.

\bibitem{bonawitz2019towards}
K.~Bonawitz, H.~Eichner, W.~Grieskamp, D.~Huba, A.~Ingerman, V.~Ivanov,
  C.~Kiddon, J.~Konecny, S.~Mazzocchi, H.~B. McMahan \emph{et~al.}, ``Towards
  federated learning at scale: {System} design,'' \emph{arXiv preprint
  arXiv:1902.01046}, 2019.

\bibitem{du2019high}
Y.~Du, S.~Yang, and K.~Huang, ``High-dimensional stochastic gradient
  quantization for communication-efficient edge learning,'' \emph{arXiv
  preprint arXiv:1910.03865}, 2019.

\bibitem{aji2017sparse}
A.~F. Aji and K.~Heafield, ``Sparse communication for distributed gradient
  descent,'' \emph{arXiv preprint arXiv:1704.05021}, 2017.

\bibitem{abad2019hierarchical}
M.~S.~H. Abad, E.~Ozfatura, D.~Gunduz, and O.~Ercetin, ``Hierarchical federated
  learning across heterogeneous cellular networks,'' \emph{arXiv preprint
  arXiv:1909.02362}, 2019.

\bibitem{lin2017deep}
Y.~Lin, S.~Han, H.~Mao, Y.~Wang, and W.~J. Dally, ``Deep gradient compression:
  {Reducing} the communication bandwidth for distributed training,''
  \emph{arXiv preprint arXiv:1712.01887}, 2017.

\bibitem{yang2018federated}
K.~Yang, T.~Jiang, Y.~Shi, and Z.~Ding, ``Federated learning via over-the-air
  computation,'' \emph{arXiv preprint arXiv:1812.11750}, 2018.

\bibitem{zhubroad2019band}
G.~{Zhu}, Y.~{Wang}, and K.~{Huang}, ``Broadband analog aggregation for
  low-latency federated edge learning,'' \emph{IEEE Trans. Wireless Commun.},
  pp. 1--1, 2019.

\bibitem{amiri2019machine}
M.~M. Amiri and D.~Gunduz, ``Machine learning at the wireless edge:
  {Distributed} stochastic gradient descent over-the-air,'' \emph{arXiv
  preprint arXiv:1901.00844}, 2019.

\bibitem{wang2019adaptive}
S.~Wang, T.~Tuor, T.~Salonidis, K.~K. Leung, C.~Makaya, T.~He, and K.~Chan,
  ``Adaptive federated learning in resource constrained edge computing
  systems,'' \emph{IEEE J. Sel. Areas Commun.}, vol.~37, no.~6, pp. 1205--1221,
  Jun. 2019.

\bibitem{kamp2018efficient}
M.~Kamp, L.~Adilova, J.~Sicking, F.~H{\"u}ger, P.~Schlicht, T.~Wirtz, and
  S.~Wrobel, ``Efficient decentralized deep learning by dynamic model
  averaging,'' in \emph{Proc. Joint European Conf. Machine Learning Knowledge
  Discovery Databases}.\hskip 1em plus 0.5em minus 0.4em\relax Springer, 2018,
  pp. 393--409.

\bibitem{chen2016revisiting}
J.~Chen, X.~Pan, R.~Monga, S.~Bengio, and R.~Jozefowicz, ``Revisiting
  distributed synchronous {SGD},'' \emph{arXiv preprint arXiv:1604.00981},
  2016.

\bibitem{nishio2019client}
T.~{Nishio} and R.~{Yonetani}, ``Client selection for federated learning with
  heterogeneous resources in mobile edge,'' in \emph{Proc. IEEE Int. Conf.
  Commun.(ICC)}, Shanghai, China, May 2019, pp. 1--7.

\bibitem{yang2019age}
H.~H. Yang, A.~Arafa, T.~Q. Quek, and H.~V. Poor, ``Age-based scheduling policy
  for federated learning in mobile edge networks,'' \emph{arXiv preprint
  arXiv:1910.14648}, 2019.

\bibitem{chen2018lag}
T.~Chen, G.~Giannakis, T.~Sun, and W.~Yin, ``{LAG: Lazily aggregated gradient
  for communication-efficient distributed learning},'' in \emph{Proc. Advances
  Neural Inf. Process. Systems}, 2018, pp. 5050--5060.

\bibitem{chen2019joint}
M.~Chen, Z.~Yang, W.~Saad, C.~Yin, H.~V. Poor, and S.~Cui, ``A joint learning
  and communications framework for federated learning over wireless networks,''
  \emph{arXiv preprint arXiv:1909.07972}, 2019.

\bibitem{zeng2019energy}
Q.~Zeng, Y.~Du, K.~K. Leung, and K.~Huang, ``Energy-efficient radio resource
  allocation for federated edge learning,'' \emph{arXiv preprint
  arXiv:1907.06040}, 2019.

\bibitem{ren2019accelerating}
J.~Ren, G.~Yu, and G.~Ding, ``Accelerating {DNN} training in wireless federated
  edge learning system,'' \emph{arXiv preprint arXiv:1905.09712}, 2019.

\bibitem{shi2019device}
W.~Shi, S.~Zhou, and Z.~Niu, ``Device scheduling with fast convergence for
  wireless federated learning,'' \emph{arXiv preprint arXiv:1911.00856}, 2019.

\bibitem{ma2017distributed}
C.~Ma, J.~Kone{\v{c}}n{\`y}, M.~Jaggi, V.~Smith, M.~I. Jordan,
  P.~Richt{\'a}rik, and M.~Tak{\'a}{\v{c}}, ``Distributed optimization with
  arbitrary local solvers,'' \emph{Optimization Methods and Software}, vol.~32,
  no.~4, pp. 813--848, Feb. 2017.

\bibitem{dinh2019federated}
C.~Dinh, N.~H. Tran, M.~N. Nguyen, C.~S. Hong, W.~Bao, A.~Zomaya, and
  V.~Gramoli, ``Federated learning over wireless networks: {Convergence}
  analysis and resource allocation,'' \emph{arXiv preprint arXiv:1910.13067},
  2019.

\bibitem{liu2010distributed}
K.~{Liu} and Q.~{Zhao}, ``Distributed learning in multi-armed bandit with
  multiple players,'' \emph{IEEE Trans. Signal Process.}, vol.~58, no.~11, pp.
  5667--5681, Nov. 2010.

\bibitem{Ali2020sleeping}
S.~{Ali}, A.~{Ferdowsi}, W.~{Saad}, N.~{Rajatheva}, and J.~{Haapola},
  ``Sleeping multi-armed bandit learning for fast uplink grant allocation in
  machine type communications,'' \emph{IEEE Trans. Commun.}, p. early access,
  2020.

\bibitem{bonnefoi2017multi}
R.~Bonnefoi, L.~Besson, C.~Moy, E.~Kaufmann, and J.~Palicot, ``Multi-armed
  bandit learning in iot networks: Learning helps even in non-stationary
  settings,'' in \emph{Proc. Int. Conf. Cognitive Radio Oriented Wireless
  Networks}.\hskip 1em plus 0.5em minus 0.4em\relax Lisbon, Portugal: Springer,
  Sep. 2017, pp. 173--185.

\bibitem{auer2002finite}
P.~Auer, N.~Cesa-Bianchi, and P.~Fischer, ``Finite-time analysis of the
  multiarmed bandit problem,'' \emph{Machine learning}, vol.~47, no. 2-3, pp.
  235--256, 2002.

\bibitem{haddadpour2019convergence}
F.~Haddadpour and M.~Mahdavi, ``On the convergence of local descent methods in
  federated learning,'' \emph{arXiv preprint arXiv:1910.14425}, 2019.

\bibitem{li2019convergence}
X.~Li, K.~Huang, W.~Yang, S.~Wang, and Z.~Zhang, ``On the convergence of fedavg
  on {non-IID} data,'' \emph{arXiv preprint arXiv:1907.02189}, 2019.

\bibitem{ahmad2009multi}
S.~H.~A. Ahmad and M.~Liu, ``Multi-channel opportunistic access: A case of
  restless bandits with multiple plays,'' in \emph{Proc. Annual Allerton Conf.
  Commun., Control, Computing (Allerton)}.\hskip 1em plus 0.5em minus
  0.4em\relax Monticello, IL, USA: IEEE, Oct. 2009, pp. 1361--1368.

\bibitem{whittle1988restless}
P.~Whittle, ``Restless bandits: {Activity} allocation in a changing world,''
  \emph{J. Applied Probability}, vol.~25, no.~A, pp. 287--298, 1988.

\bibitem{li2019combinatorial}
F.~{Li}, J.~{Liu}, and B.~{Ji}, ``Combinatorial sleeping bandits with fairness
  constraints,'' in \emph{Proc. IEEE Conf. Computer Commun. (INFOCOM)}, Paris,
  France, Apr. 2019, pp. 1702--1710.

\bibitem{stich2018local}
S.~U. Stich, ``Local {SGD} converges fast and communicates little,''
  \emph{arXiv preprint arXiv:1805.09767}, 2018.

\bibitem{chen2020convergence}
M.~Chen, H.~V. Poor, W.~Saad, and S.~Cui, ``Convergence time optimization for
  federated learning over wireless networks,'' \emph{arXiv preprint
  arXiv:2001.07845}, 2020.

\bibitem{lai1985asymptotically}
T.~L. Lai and H.~Robbins, ``Asymptotically efficient adaptive allocation
  rules,'' \emph{Advances in Applied Mathematics}, vol.~6, no.~1, pp. 4--22,
  1985.

\bibitem{bubeck2012regret}
S.~Bubeck, N.~Cesa-Bianchi \emph{et~al.}, ``Regret analysis of stochastic and
  nonstochastic multi-armed bandit problems,'' \emph{Foundations and
  Trends{\textregistered} in Machine Learning}, vol.~5, no.~1, pp. 1--122,
  2012.

\bibitem{gai2012combinatorial}
Y.~{Gai}, B.~{Krishnamachari}, and R.~{Jain}, ``Combinatorial network
  optimization with unknown variables: {Multi-Armed} bandits with linear
  rewards and individual observations,'' \emph{IEEE/ACM Trans. Networking},
  vol.~20, no.~5, pp. 1466--1478, Oct. 2012.

\bibitem{gupta2019multiarmed}
S.~Gupta, S.~Chaudhari, G.~Joshi, and O.~Ya\v{g}an, ``Multi-armed bandits with
  correlated arms,'' \emph{arXiv preprint arXiv:1911.03959}, 2019.

\bibitem{xie2018zeno}
C.~Xie, O.~Koyejo, and I.~Gupta, ``Zeno: Distributed stochastic gradient
  descent with suspicion-based fault-tolerance,'' \emph{arXiv preprint
  arXiv:1805.10032}, 2018.

\bibitem{neely2010stochastic}
M.~J. Neely, ``Stochastic network optimization with application to
  communication and queueing systems,'' \emph{Synthesis Lectures Commun.
  Networks}, vol.~3, no.~1, pp. 1--211, 2010.

\bibitem{hsu2018intergrating}
W.~{Hsu}, J.~{Xu}, X.~{Lin}, and M.~R. {Bell}, ``Integrating online learning
  and adaptive control in queueing systems with uncertain payoffs,'' in
  \emph{Inf. Theory Appli. Workshop (ITA)}, San Diego, CA, USA, Feb. 2018, pp.
  1--9.

\bibitem{chen2020efficient}
M.~Chen, B.~Mao, and T.~Ma, ``Efficient and robust asynchronous federated
  learning with stragglers,'' in \emph{Proc. Int. Conf. Learning
  Representations (ICLR)}, 2020.

\bibitem{yang2019scheduling}
H.~H. {Yang}, Z.~{Liu}, T.~Q.~S. {Quek}, and H.~V. {Poor}, ``Scheduling
  policies for federated learning in wireless networks,'' \emph{IEEE Trans.
  Commun.}, pp. 1--1, 2019.

\bibitem{pollard2012convergence}
D.~Pollard, \emph{Convergence of stochastic processes}.\hskip 1em plus 0.5em
  minus 0.4em\relax Springer Science \& Business Media, 2012.

\end{thebibliography}

\end{document}